\documentclass[manuscript]{aastex}
\usepackage{threeparttable,booktabs}

\slugcomment{For submission to the Astrophysical Journal}

\shorttitle{Multiple Bipolar Molecular Outflows from L1551~IRS55}

\shortauthors{Wu et al.}

\begin{document}

\title{Multiple Bipolar Molecular Outflows from the L1551~IRS5 Protostellar System}

\author{Po-Feng Wu,\altaffilmark{1,2} Shigehisa Takakuwa,\altaffilmark{2} and Jeremy Lim\altaffilmark{2}}

\altaffiltext{1}{Institute of Astrophysics, National Taiwan University, No. 1, Sec. 4, Roosevelt Road, Taipei 106, Taiwan}
\altaffiltext{2}{Academia Sinica Institute of Astronomy and Astrophysics, P.O. Box 23-141, Taipei, 106, Taiwan}

\begin{abstract}
The multiple protostellar system L1551~IRS5 exhibits a large-scale bipolar molecular outflow that spans $\sim$1.5~pc on both the NE (redshifted) and SW (blueshifted) sides of the system.  We have studied this outflow within $\sim$4000~AU of its driving source(s) with the SubMillimeter Array.  Our CO(2--1) image at $\sim$4\arcsec\ ($\sim$560~AU) resolution reveals three distinct components: 1) an X-shaped structure spanning $\sim$20\arcsec\ from center with a similar symmetry axis and velocity pattern as the large-scale outflow; 2) an S-shaped structure spanning $\sim$10\arcsec\ from center also with a similar symmetry axis but opposite velocity pattern to the large-scale outflow; and 3) a compact central component spanning $\sim$1.4\arcsec\ from center again with a similar symmetry axis and velocity pattern as the large-scale outflow.  The X-shaped component likely comprises the limb-brightened walls of a cone-shaped cavity excavated by the outflows from the two main protostellar components.  The compact central component likely comprises material within this cavity newly entrained by one or both outflows from the two main protostellar components.  The S-shaped component mostly likely comprises a precessing outflow with its symmetry axis inclined in the opposite sense to the plane of the sky than the other two components, taking the S-shaped component out of the cone-shaped cavity along most if not all of its entire length.  This outflow may be driven by a recently reported candidate third protostellar component in L1551~IRS5, whose circumstellar disk is misaligned relative to the two main protostellar components.  Gravitational interactions between this protostellar component and its likely more massive northern (and perhaps also southern) neighbor(s) may be causing the circumstellar disk and hence outflow of this component to precess.
\end{abstract}


\section{Introduction}
Most near-solar-mass stars are born as members of binary or multiple systems \citep{duq91,mat94}.  The formation of such systems, although the preferred mode of star formation at least for near-solar-mass stars, is poorly understood.
The main competing theoretical models are fragmentation of dense molecular condensations to form systems comprising multiple protostars, or capture of originally single protostars still embedded in their individual condensations to form multiple systems \citep[e.g., review by][]{toh02}.

The most direct way of distinguishing between these two possibilities is to compare for a given system the properties of its individual protostars (i.e., orientation of their circumstellar disks and their orbital motion) with the properties of its surrounding molecular condensation (i.e., orientation and rotation).  Such a study of the low-mass protostellar system L1551~IRS5 provides the most direct evidence yet that its two main protostellar components formed via the fragmentation of their surrounding parent molecular condensation \citep{lim06}.  For many systems (especially those with closely separated components), however, the properties of their individual protostellar components are difficult to measure until the Extended Very Large Array (EVLA) or Atacama Large Millimeter and subMillimeter Array (ALMA) is completed.  Indeed, at the present time, even the number of protostellar components present in a given system is not always clear.

Bipolar molecular outflows, found ubiquitously around protostars, are more easily accessible, and can be used to indirectly probe the properties of multiple protostellar systems.  For example, such outflows can reveal the presence of currently undetectable or spatially unresolved protostellar components through their individual outflows. 
Outflows that wind in space with a helical pattern may have driving sources whose circumstellar disks and therefore jets are precessing due to gravitational interactions with closely-separated protostellar companions.  The degree of collimation and/or dynamical age of the outflow, as well as the opening angle of the outflow cavity at its base, can provide a guide to the evolutionary stage of the driving source.

Here, we study the bipolar molecular outflow associated with L1551~IRS5 at a high angular resolution with an interferometric array.  This system was first found as an infrared 
source by \citet{sto76} in the nearby (distance $\sim$140 pc) L1551 
molecular cloud located in the constellation Taurus.  L1551~IRS5 was the first protostellar source from which a bipolar molecular outflow, detected in single-dish observations of the CO(1--0) and CO(2--1) lines, was recognized \citep{sne80}.  Subsequent single-dish observations in CO(1--0), CO(2--1), and CO(3--2) at higher angular resolutions have revealed the overall morphology and kinematics of this and a number of other outflows in the L1551 star-forming region on parsec scales \citep{uch87,bac94,mo87,mo88,mo06,sto06}.  The most prominent of these outflows is centered on L1551~IRS5, and extends $\sim$1.5 pc towards the NE (redshifted lobe) and SW (blueshifted lobe) directions at a position angle of $\sim$$50\degr$.  Each outflow lobe has a U-shaped structure, together with knotty components having distinct velocities.  The other outflows detected in the L1551 cloud have been associated with other driving sources except for a collimated redshifted outflow with one end near L1551~IRS5 and extending $\sim$2 pc to the west, the origin of which is not known.

Apart from the large-scale bipolar molecular outflow, L1551~IRS5 is surrounded by a relatively compact molecular condensation (radius of $\sim$1200~AU in C$^{18}$O) that has been best studied by \citet{mom98}.  This condensation is elongated in the NW-SE direction, perpendicular to the major axis of the large-scale bipolar molecular outflow.  The measured kinematics indicate a  flattened envelope that is rotating and contracting, as also suggested in earlier observations by \citet{oha96} and \citet{sai96}.  This condensation has been observed in CS (7--6), which traces warmer and denser gas closer to the central protostellar system, by \citet{tak04}.  In the inner region of the envelope, the kinematics is dominated by rotation.

The first indication that L1551~IRS5 may be a binary system was reported by \citet{bie85}, who showed that this system comprised two sources separated by $\sim$0.3\arcsec\ ($\sim$42~AU) in the north-south direction at a wavelength of 2~cm.  Subsequent studies resolved these two sources into twin ionized jets closely aligned with the large-scale bipolar molecular outflow \citep{rod03b}.  \citet{loo97} showed from high angular-resolution ($\sim$0\farcs5) observations in the 2.7-mm continuum that L1551~IRS5 contains two circumstellar dust disks surrounded by a circumbinary dust disk and/or extended envelope.  \citet{rod98} spatially resolved the two circumstellar dust disks (along their major axes) for the first time at 7~mm, and showed that they are centered on the above-mentioned twin ionized jets.  \citet[][]{rod03a} showed that these two protostellar components were (likely) orbiting each other in a clockwise direction.  \citet{lim06} spatially resolved the two circumstellar disks along both their major and minor axes, and showed that these disks are aligned with each other as well their surrounding rotating and flattened molecular condensation (pseudodisk).  They also showed that the two components were orbiting each other in the same direction as the rotation of the pseudodisk.  All these properties are consistent with the notion that the two protostellar components formed as a result of fragmentation in the central region of their surrounding pseudodisk.  Finally, \citet{lim06} found a candidate third protostellar component located (in projection) near the northern protostellar component.  The circumstellar disk of this third component is misaligned relative to the circumstellar disks of the two main components as well as their surrounding pseudodisk.

We report in this paper high angular resolution ($\sim$4\arcsec) observations in CO(2--1) centered on the base of the large-scale bipolar molecular outflow from L1551~IRS5.  There is, to the best of our knowledge, only one other published interferometric image of this molecular outflow in CO, that presented by \citet{bar93} in CO(1--0) who studied relatively extended ($\gtrsim 1\arcmin$) features in the blue outflow lobe.  In our study, we investigate the structure and kinematics of the outflow components detected within $\sim$30\arcsec\ of L1551~IRS5, their connection with the large-scale bipolar molecular outflow imaged with single-dish telescopes, and the likely nature of their driving sources.  

Readers interested in how the observations were conducted and data reduced should now proceed to $\S$2.  Those interested only in the results of the CO(2--1) and simultaneous 1.3~mm continuum measurements can skip ahead to $\S$3.  In $\S$4, we show that our CO(2--1) maps likely traces three separate outflow components: the limb-brightened walls of an outflow cavity, material in the cavity entrained by one or both jets from the two main protostellar components, and a candidate third precessing bipolar outflow.  Readers interested only in a concise summary of our results and interpretation can proceed directly to $\S$5.

\section{Observations and Data Reduction}
Observations of L1551~IRS5 in CO(2--1) and 1.3 mm continuum were carried out using the SMA\footnote{The Submillimeter Array (SMA) is a joint project between the Smithsonian Astrophysical Observatory and the Academia Sinica Institute of Astronomy and Astrophysics and is funded by the Smithsonian Institution and the Academia Sinica.} in its compact configuration on 2003 December 7.  A description of the SMA can be found in \citet{ho04}.  The parameters of the observations are summarized in Table~1.  Seven of the eight antennas of the SMA were available for the observations.  We used the stronger quasar 0423-013, lying 19.6$\degr$ from L1551~IRS5, for amplitude calibration, and the weaker quasar 3C~120, lying 12.8$\degr$ from L1551~IRS5, for phase calibration.  The flux density of 0423-013 was 9.4~Jy and that of 3C~120 was 3.3~Jy as measured with respect to Uranus, which served as the absolute flux calibrator.  

The SMA has a double-sideband receiver in the 230~GHz (1.3~mm) band with a bandwidth of 2 GHz in each sideband.  The lower sideband was tuned to a central frequency of 231.3~GHz, and the upper sideband 241.3~GHz.  The signal in each sideband is fed into a correlator that distributes this signal to twenty-four spectral windows (''chunks'' in the SMA nomenclature).  Six of the chunks were divided into 512 channels, which in the CO(2--1) line results in a velocity resolution of 0.265 km s$^{-1}$ over a velocity range of 135 km s$^{-1}$.  The remaining chunks were divided into 128 channels, and following calibration vector-averaged to make a single continuum channel at 1.3~mm with a total bandwidth of 4.0~GHz.  The minimum projected baseline in our observation was $\sim$8~k$\lambda$, allowing us to recover $\sim$$30\%$ of the flux from the largest structure visible in our CO(2--1) maps of $\sim$15\arcsec\ \citep[e.g., see][]{wil94}. 

We calibrated the raw visibility data using MIR, which is an IDL-based data reduction package adopted for the SMA from the MMA software package originally developed for OVRO \citep{sco93}.  We adopted antenna-based calibration, which provides the best sensitivity, as there were no significant baseline-based errors in our data.  The calibrated visibility data were Fourier-transformed to produce DIRTY images, and the point spread function of the telescope deconvolved (CLEANed) from these images using MIRIAD \citep{sau95} to produce the final maps.  A ROBUST parameter of 0.5 was adopted, which provided the best compromise between sensitivity and angular resolution.

To make the CO(2--1) channel maps, we first subtracted the continuum 
emission from the visibility data as derived from a fit to the line-free channels.  Unlike in the continuum where the emission is centrally concentrated, in CO(2--1) the emission in many channels is distributed in a complex manner over an extended region resulting in complicated sidelobe patterns.  To CLEAN the channel maps, we first selected (i.e., place boxes enclosing) features that we believed to be real.  Through trial and error, we converged on features that changed smoothly in structure between neighboring channels, as would be expected for real features.  An examination of the residual maps (i.e., after the candidate real features and their sidelobes had been removed) revealed no systematic pattern (i.e., only random noise), indicating that all the real feature had indeed been correctly recovered.  Thus, we believe that any artificial components incorrectly selected for CLEANing are much weaker than the real components in our maps.

\section{Results}

\subsection{1.3~mm Continuum}
Figure~1 shows our 1.3-mm continuum map of L1551~IRS5 at an angular resolution of $4\farcs66 \times 2\farcs33$.  The continuum emission peaks at the location of the protostellar system, as seen in images at comparable angular resolution at 3~mm \citep{loo97} and 0.8~mm \citep{tak04}.
We fitted a two-dimensional Gaussian to the continuum source to derive a total flux density of 1.20$\pm$0.01~Jy and a deconvolved size at full-width half-maximum (FWHM) of $(1\farcs5 \pm 0\farcs1) \times (0\farcs8 \pm 0\farcs2)$ along a position angle of $146\degr \pm 8\degr$.  The size of the continuum source is much larger than the separation of the two main protostellar components ($\sim$0.3$\arcsec$), implying that at least a part of the emission arises from the surrounding envelope.  \citet{loo97} found that, at 3~mm, about half of the continuum emission arises from the circumstellar dust disks of the two main protostellar components (their observation did not have sufficient angular resolution to spatially separate the candidate third protostellar component), and about half from the surrounding dust envelope.  The continuum source detected here has an orientation similar to but size much smaller than the pseudodisk traced in C$^{18}$O by \citet{mom98}, suggesting that the extended part of the continuum emission traces the inner regions of the pseudodisk.

Both free-free emission from ionized jets and thermal emission from dust can contribute to the continuum emission.  In L1551~IRS5, as in many protostellar systems, the ionized jets dominate the emission at cm wavelength, whereas dust contributes an increasingly larger fraction of the emission towards shorter wavelengths.  To assess the relative contribution from each component at a given wavelength, we have collected all relevant measurements of the continuum emission for L1551~IRS5 at centimeter to submillimeter wavelength as listed in Table~2.  Figure~2 shows the spectral energy distribution (SED) of the continuum emission based on the tabulated data.  We modeled the SED as the sum of two power-law components with each component having $F_\nu \propto \nu^{\alpha}$, where $F_\nu$ is the intensity at a given frequency $\nu$ and $\alpha$ the power-law index.  A chi-squared fit to the data gives one component with $\alpha = -0.1 \pm 0.1$ that dominates at low frequencies (short-dashed line in Fig.~2), consistent with optically-thin free-free emission.  The other component has $\alpha = 3.0 \pm 0.1$ that dominates at high frequencies (long-dashed line in Fig.~2), consistent with thermal emission from dust.  As can be seen, the contribution from free-free emission to the continuum at 1.3 mm is only $\sim$$0.1\%$. 

Assuming that the dust temperature ($T_d$) is uniform, we can estimate the total dust mass of the continuum source from the relationship 
\begin{equation}
M = \frac{F_{\nu} \ d^{2}} {B_{\nu}(T_d) \ \kappa_{\nu}} \ \ ,
\end{equation}
where $d$ is the distance to the source, $B_{\nu}$ is the blackbody function, $T_d$ the dust temperature and $\kappa_{\nu}$ the dust mass opacity coefficient given by $\rm \kappa_{\nu} = 0.1 \ (\nu/10^{12} \ Hz)^{\beta} \ cm^2 \ g^{-1}$ \citep{bec91}.  The quantity $\beta$ is related to the spectral index, $\alpha$, of the dust continuum emission by the relationship $\alpha$ = 2 + $\beta$.  Assuming $T_d$ = 47 K \citep{mo94}, the dust mass inferred from Equation~(1) is $0.06 \pm 0.01 {\rm \ M_\sun}$.
This value is in reasonable agreement with those derived at 2.7~mm that are sensitive to the same size scales \citep{oha91,mom98}.

\subsection{CO (2--1)}
Figure~3 shows our CO (2--1) channel maps of L1551~IRS5.  For presentation purposes, none of the maps displayed in this manuscript have been corrected for the primary beam response of the SMA; the quantities we computed were, of course, derived from maps corrected for the primary beam response.  Emission is detectable over the velocity range 0.2--$10.5 {\rm \ km \ s^{-1}}$ measured with respect to the local standard of rest ($V_{\rm LSR}$).  We estimate based on the symmetry of the features observed (see below) a systemic velocity of $\sim$6.5 km s$^{-1}$ for the L1551~IRS5 protostellar system, comparable to previous estimates \citep{fri02}.  The absence of emission within $\sim$$0.5 {\rm \ km \ s^{-1}}$ of the systemic velocity is likely caused by extended emission (mostly from the ambient molecular cloud) that is resolved out by the interferometer.

\subsubsection{X-shaped structure}
Figure~4 shows the integrated CO(2--1) intensity map at blueshifted velocities spanning 0.2--$5.5 {\rm \ km \ s^{-1}}$ (blue contours) and redshifted velocities spanning 7.4--$10.5 {\rm \ km \ s^{-1}}$ (red contours) separately.  To make this map, we first applied Hanning smoothing (a triangular function of width five channels at zero intensity) to the channel maps, and then excluded all features in the channel maps where the intensity in the corresponding smoothed maps is less than 1.5$\sigma$.  The most prominent feature in this map is an X-shaped structure (as indicated by the diagonal lines) that has its blueshifted pair of arms lying west and its redshifted pair of arms lying east of the protostellar system.  In the channel maps of Figure~3, the blueshifted part of this X-shaped structure can be most clearly seen at velocities spanning 2.3--$5.2 {\rm \ km \ s^{-1}}$ and the redshifted part 7.6--$10.5 {\rm \ km \ s^{-1}}$.  Both pairs of arms have an opening angle of $\sim$$90\degr$, and extend $\sim$20\arcsec--30\arcsec\ ($\sim$3000--4000~AU) from center.  They have a symmetry axis along the NE--SW direction at a position angle of about $70\degr$, which is close to the orientation of the major axes of the ionized jets as well as the large-scale bipolar molecular outflow.  Furthermore, the X-shaped structure exhibits the same velocity pattern (i.e., redshifted to the E and blueshifted to the W) as the large-scale bipolar molecular outflow.

In both the channel maps of Figure~3 and integrated-intensity map of Figure~4, the blueshifted NW arm can be seen to be much broader than the other three arms of the X-shaped structure.  A close examination of the channel maps reveals that the NW arm comprises several distinct and relatively compact (bullet-like) spatial-kinematic components.  For example, the broadest portion of the NW arm at about 10$\arcsec$ from center (see Fig.~4) is dominated by a pair of comparatively high-velocity bullets that can be best seen in the channel maps (Fig.~3) at velocities of 2.1$\rm \ km \ s^{-1}$--3.4$\rm \ km \ s^{-1}$.  The tip of the NW arm is dominated by two lower-velocity bullets that can be best seen in the channel maps at 3.7$\rm \ km \ s^{-1}$--4.7$\rm \ km \ s^{-1}$.  On the other hand, the two redshifted arms each comprise only one spatial-kinematic component (see below).  In this way, the properties of the X-shaped component seem to mirror that of the large-scale bipolar molecular outflow.  Several distinct spatial-kinematic components have been seen in single-dish CO(2--1) observations spaced at semi-regular intervals along the blueshifted outflow lobe of L1551~IRS5 \citep{bac94}.  Curiously, no such distinct components are present in the same observations of the redshifted outflow lobe.

Figure~5 shows position-velocity (PV-) diagrams along each opposing set of arms as indicated by the diagonal lines in Figure~4.  The straight lines drawn in Figure~5 are not meant to guide the eye, but instead are our model fits to the PV-diagrams as we shall describe later in $\S4.1$.  The PV-diagrams along the NE (Fig.~5(a), upper right quadrant) and SE (Fig.~5(b), upper right quadrant) arms look very similar, with each having a velocity at peak intensity that increases with radial distance (i.e., Hubble-like expansion) out to $\sim$20\arcsec\ from center.  The SW arm (Fig.~5(a), lower left quadrant) is only weakly detected in the PV-diagram, but nevertheless can be seen to have a velocity gradient comparable to the NE and SE arms.  On the other hand, the PV-diagram along the NW arm (Fig.~5(b), lower left quadrant), shows at least two distinct velocity components.  The component labeled NW-a increases comparatively quickly in velocity with radial distance out to $\sim$10$\arcsec$ from center.  The component labeled NW-b increases more slowly in velocity with radial distance out to $\sim$30$\arcsec$ from center.  NW-a corresponds to the pair of higher-velocity bullets, and NW-b to the pair of lower-velocity bullets, mentioned above.  Neither of these components have a velocity gradient that matches the comparable velocity gradients of the other three arms of the X-shaped component.

\subsubsection{S-shaped structure}
The next most prominent feature in the integrated-intensity map of Figure~4 is located between the blueshifted and redshifted arms of the X-shaped structure and extends to $\sim$10$\arcsec$ from center.  The outer portion of this feature on the NE side appears to twist north and that on the SW side twist south, creating a S-shaped structure.  Its symmetry axis is oriented at a position angle of about 55$\degr$, approximately aligned with the symmetry axis of the X-shaped component as well as major axis of the large-scale bipolar molecular outflow.

The S-shaped structure exhibits blueshifted velocities NE of center, and redshifted velocities SW of center.  The velocity pattern of the S-shaped structure is therefore opposite to that of both the X-shaped structure and large-scale bipolar molecular outflow.  In the channel maps of Figure 3, the blueshifted part of the S-shaped structure can be seen most clearly at velocities spanning 3.4--$5.5 {\rm \ km \ s^{-1}}$ and the redshifted part 7.4--$8.2 {\rm \ km \ s^{-1}}$.  The line-of-sight velocity of the S-shaped structure therefore does not reach velocities as high (far from $V_{\rm sys}$) as the X-shaped structure.  As described in $\S\ref{Precessing Outflow}$, this structure exhibits an approximately sinusoidal variation in its radial velocity along its length.

\subsubsection{Compact central component}\label{a compact central component}
Finally, a compact component centered on the location of L1551~IRS5 can be seen in the channel maps of Figure~3 extending to higher blueshifted and redshifted velocities than either the X- or S-shaped components.  It is the only detectable or dominant component visible at blueshifted velocities spanning 0.2--$3.4 {\rm \ km \ s^{-1}}$ and redshifted velocities spanning 9.2--$10.5 {\rm \ km \ s^{-1}}$.  In the integrated-intensity map of Figure~4, although difficult to visually separate from the base of the X- or S-shaped structures, the compact central component comprises the closest redshifted and blueshifted intensity peaks straddling the L1551~IRS5 protostellar system.

To better extract the properties of the compact central component, we have fitted a two-dimensional gaussian structure to this component at each velocity where it is detectable in the channel maps.  Figure~6 shows the integrated CO(2--1) intensity map made from the derived fits with the blueshifted and redshifted emission plotted separately.  The centroids of the redshifted and blueshifted emissions are displaced by $\sim$1.4\arcsec\ ($\sim$200~AU) NE and SW from center, respectively, at a position angle of about 65\degr.  The kinematic axis of this component is therefore closely aligned with the symmetry/major axis of, and has the same velocity pattern as, the X-shaped component and large-scale bipolar molecular outflows.

\subsubsection{Base of X-shaped and S-shaped components}
To better study the structure of the X- and S-shaped components near their base, we have subtracted the compact central component from the channel maps using the above-mentioned fits to this component ($\S\ref{a compact central component}$).  Figure~7 shows the integrated CO(2--1) intensity maps made before (upper row) and after (lower row) the compact central component was subtracted, separated into two different velocity regimes.  The higher velocity regime (Fig.~7, left column) spans 0.2--$3.4 {\rm \ km \ s^{-1}}$ (i.e., $< V_{\rm LSR}-3.1 {\rm \ km \ s^{-1}}$) at blueshifted velocities (blue contours) and 8.9--$10.5 {\rm \ km \ s^{-1}}$ ($> V_{\rm LSR}+2.4 {\rm \ km \ s^{-1}}$) at redshifted velocities (red contours).  The lower velocity regime (Fig.~7, right column) spans 3.7--$5.5 {\rm \ km \ s^{-1}}$ ($> V_{\rm LSR}-2.8 {\rm \ km \ s^{-1}}$) at blueshifted velocities (blue contours) and 7.4--$8.7 {\rm \ km \ s^{-1}}$ ($< V_{\rm LSR}+2.2 {\rm \ km \ s^{-1}}$) at redshifted velocities (red contours).  Like before, to make these maps we first applied Hanning smoothing (a triangular function of width five channels at zero intensity) to the channel maps, and then excluded all features in the channel maps where the intensity in the corresponding smoothed maps is less than 1.5$\sigma$.  

The S-shaped component, which as mentioned above does not reach velocities as high as the X-shaped component, is barely detectable in the higher velocity regime.  In this regime, the compact central component can be well separated from the X-shaped component as can be seen by comparing the map made before and after subtraction.  In the lower velocity regime, all three components are detectable.  The compact central component dominates the emission close to the protostellar system as can be seen by comparing the map made before and after subtraction.

\subsubsection{Physical parameters}
We have estimated physical parameters for all three components separately after correcting our maps for the primary beam response of the SMA antennas.  For simplicity, we assumed that the observed CO(2--1) emission is optically thin, 
in local thermal equilibrium (LTE), and has an excitation temperature of 
10~K as derived by \citet{sto06} based on the intensity ratio between CO(3--2) and CO(1--0).  We also assumed a H$_2$-to-CO abundance ratio of $10^4$.  We adopted an inclination of $30\degr$ with respect to the plane of the sky for all three components, such that their major axes are orthogonal to the equatorial plane of the circumstellar disks of the two main protostellar components and their surrounding flattened condensation \citep{lim06}.  As we shall discuss later (\S4.3), the adopted inclination is probably appropriate for the X-shaped and compact central components, and close to median value in the allowed range from our model fit for the S-shaped component (but with the NE side the near side, opposite to the X-shaped and compact central components).  

In Table~3, we tabulate for each component its estimated mass $M$, outflow momentum $P = \sum_{v} M(v) \ v$, kinetic energy $E = \frac{1}{2} \sum_{v} M(v) \ v^2$, mass loss rate $\dot{M} = \sum_v M(v) / \tau(v)$, and dynamical age corresponding to the maximum derived dynamical time $\tau(v)$ = ${v}/{r}$, where $r$ is the distance from the driving source to the emission peak of the outflowing gas at a given velocity $v$.  Note that all these values except for dynamical age are lower limits, in part because we have assumed that the CO(2-1) emission is optically thin, and in part because emission close to the systemic velocity was entirely resolved out.  The X-shaped component, which is the most obvious component in our map, dominates in mass, momentum, and energy.  It also has the oldest dynamical age.  The mass-loss rate is highest for the compact central component, which also has the youngest dynamical age.

\section{Discussion}
Our CO(2--1) map at an angular resolution of $2\farcs3 \times 4\farcs7$ reveals, for the first time, three distinct bipolar molecular outflow components in the vicinity ($\lesssim 4000$~AU) of the L1551~IRS5 protostellar system.  In the following, we shall discuss the likely nature of each of these components as well as their driving sources.

\subsection{X-Shaped Component}
The X-shaped component has a symmetry axis closely aligned with and exhibits the same velocity pattern as the large-scale bipolar molecular outflow.  It has the largest mass, momentum, and energy of the three outflow components that we detected in the vicinity of the L1551~IRS5 protostellar system.  This X-shaped component resembles similar structures seen around a number of other protostars such as IRS1 in Barnard~5 \citep{vel98}, IRAS 20582+7724 in L1228 \citep{arc04}, and IRAS~5295+1247 in L1528B \citep{arc05}.  In all these cases, the X-shaped structure observed is thought to comprise the limb-brightened walls of a cone-shaped cavity entrained and excavated by outflows from the central protostars.  Indeed, scattered light from a cone-shaped cavity can be seen in the near-IR on the SW side of L1551~IRS5 \citep{hod94} with an opening angle comparable to that measured for the (blueshifted) arms of the X-shaped component.  (The NE side of the cavity is presumably obscured in the near-IR by material in the condensation and extended envelope.)

\citet{arc06} have conducted a $^{12}$CO survey of molecular outflows within $\lesssim$$10^4$~AU of nine protostellar (comprising a mixture of both single and multiple) systems at different evolutionary stages.  They found a clear difference in the morphology of the molecular outflow lobes between Class~0 and more evolved Class~I protostars.  Class~0 protostars exhibit relatively collimated jet-like outflows or cone-shaped lobes with opening angles $< 55\degr$, whereas Class~I protostars have outflow lobes with wider opening angles of $> 75\degr$ (with $\sim$$125\degr$ the largest opening angle seen).  L1551~IRS5 is classified (based on its Infrared SED) as a Class I protostar \citep[][]{kee90}, and so the opening angle of its outflow cavity ($\sim$$90\degr$) is comparable with that seen around other protostars at a similar evolutionary stage.

Near-IR observations in the [Fe II] line by \citet{pyo02,pyo05} have revealed two velocity components in the ionized outflow from the main northern protostellar component in L1551~IRS5.  One of these outflow components comprises a high-velocity ($\sim$$300 {\rm \ km \ s^{-1}}$) and narrowly-collimated wind, and the other a low-velocity ($\sim$$100 {\rm \ km \ s^{-1}}$) and wide-angle wind.  The latter has an opening angle of $\sim$100$\degr$, which is similar to the opening angle of the cavity seen through scattered light in the near-IR and as the X-shaped component.  The X-shaped component may therefore have formed in two steps: first the high-velocity and narrowly-collimated wind passes through and in the process entrains molecular gas \citep[e.g.,][]{sta94}, followed by a low-velocity and wide-angle wind that pushes forward and aside the molecular gas to excavate a cone-shaped cavity.  \citet{bac94} have found that the large-scale bipolar molecular outflow associated with L1551~IRS5 exhibits several successive high-velocity features along its blueshifted lobe suggestive of different ejection episodes.  The X-shaped component may therefore comprise (primarily) the base of that portion of the larger-scale outflow created in the latest ejection episode. 

We now investigate what properties a cone-shaped outflow cavity should have to reproduce both the observed structure and kinematics of the X-shaped component.  Figure~8 shows a sketch of a cone with a half-opening angle $\theta$ and its symmetry axis inclined by an angle $i$ to the plane of the sky.  As suggested by the PV-diagrams of Figure~5, we assume that the cone exhibits a Hubble-like radial expansion such that its radial velocity is given by $v_r = Cr$, where $C$ is the coefficient of the radial flow and $r$ the radial distance from the driving source.  We assume $i = 30\degr$ so that the symmetry axis is orthogonal to the circumstellar disks of the two main protostellar components \citep[see][]{lim06}, and measure from the integrated-intensity map of Figure~4 a half-opening angle of $\theta \approx 45\degr$.  The only free parameter is therefore $C$, which in the PV-diagram of Figure~5 specifies the observed velocity gradient along the arms of the X-shaped component.  We found that a satisfactory fit to all except the blueshifted NW arm can be obtained with $C = 0.18 {\rm \ km \ s^{-1} \ arcsec^{-1}}$, and plot the model fit as solid lines in the PV-diagram of Figure~5.  As mentioned earlier in $\S3.2.1$, the blueshifted NW arm exhibits two distinct velocity components in the PV-diagram of Figure~5, neither of which have velocity gradients that matches the otherwise similar velocity gradients along the other three arms. 

The acceleration constant of the Hubble-like radial expansion estimated for the X-shaped component of $C \approx 0.18 {\rm \ km \ s^{-1} \ arcsec^{-1}} \approx 250 {\rm \ km \ s^{-1} \ pc^{-1}}$ is much higher than that estimated from single-dish observations for the large-scale bipolar molecular outflow from L1551~IRS5 of $\sim$$70 {\rm \ km \ s^{-1} \ pc^{-1}}$ \citep{fri89}.  Interferometric observations of the bipolar molecular outflow from B335 reveal an even higher acceleration constant of $\sim$$490 {\rm \ km \ s^{-1} \ pc^{-1}}$ \citep{yen}.
On the other hand, the acceleration constant of the Hubble-like radial expansion of large-scale ($>$ 0.1 pc) molecular outflows in other low-mass protostars have measured values of $\sim$$65 {\rm \ km \ s^{-1} \ pc^{-1}}$ in L1157 \citep{bac01}, $\sim$$97 {\rm \ km \ s^{-1} \ pc^{-1}}$ in both L483 \citep{taf00} and BHR 71 \citep{bou97}, and $\sim$$120 {\rm \ km \ s^{-1} \ pc^{-1}}$ in IRAM 04191+1522 \citep{and99}.  All these values are lower than that measured in interferometric observations of either L1551~IRS5 or B335 that probe the base of their respective outflows, suggesting that the acceleration constant of the Hubble-like radial expansion decreases outwards along a given outflow lobe.

\subsection{Compact Central Component}
The compact central component has its major axis aligned with the symmetry axis of the X-shaped component, as well as major axis of the large-scale bipolar molecular outflow.  Furthermore, it exhibits the same velocity pattern as both these more spatially extended components.  Together with the fact that it reaches much higher blueshifted and redshifted velocities than the X-shaped component, the compact central component most likely comprises material lying along the symmetry axis of the cone-shaped cavity that has been entrained relatively recently by the jet(s) from one (or both) of the two main protostellar components.  High angular-resolution observations of the bipolar molecular outflows from other protostars also sometimes show a relatively collimated outflow lying along the symmetry axis of a cone-shaped outflow cavity \citep[e.g., HH211;][]{gue99}, with the more collimated component attributed to molecular gas entrained by a jet. 

\citet{fri98} found that the two optical jets visible on the SW side of L1551 IRS5 have velocities of $\sim$$300 {\rm \ km \ s^{-1}}$.  As mentioned in $\S4.1$, \citet{pyo02,pyo05} found that the component at higher velocities of $\sim$$300 {\rm \ km \ s^{-1}}$ in the northern optical jet is narrowly collimated.  \citet{fri98} inferred based on the observed H$\alpha$ and [S~II] line intensities a particle density of $\sim$$10^{4} {\rm \ cm^{-3}}$ in these jets, and therefore a momentum for each jet of $\sim$$7 \times 10^{-4} {\rm \ M_{\odot} \ km \ s^{-1}}$.  The combined momentum of the optical jets is therefore comparable with the momentum we inferred for the compact central component of $\sim$$1.33 \times 10^{-3} {\rm \ M_{\odot} \ km \ s^{-1}}$, demonstrating that the optical jets indeed have sufficient momentum to drive the compact central component.  At their velocities of $\sim$$300 {\rm \ km \ s^{-1}}$, the optical jets would travel the entire length of the compact central component in just $\sim$30~yrs.  The presence of newly entrained molecular gas in the outflow cavity may indicate that a significant quantity of molecular gas has recently been injected into this cavity from the surrounding infalling envelope, which tries to fill or is diverted into the cavity at or near its base.  Alternatively, or in addition, a relatively recent enhancement in the activity (i.e., mass-loss rates) of one or both optical jets have led to an enhancement in the amount of entrained material in the cavity. 

Using the Chandra X-ray Observatory, \citet{bal03} have detected a compact, but slightly resolved, X-ray source located about 0.5\arcsec--1\arcsec\ south-west of L1551~IRS5 at the base of the two optical jets.  They proposed several models for how the observed X-ray emission may be produced, all of which involve radiating plasma downstream of a shock where either: i) a wide-angle flow, launched and accelerated within a few AU of one member of the IRS 5 protobinary, is redirected and collimated into a jet; ii) one or both jets collide with a circumbinary disk; and (iii) the two jets collide.  Here, we propose that shocks produced when one or both jets slam into molecular gas diverted into the X-shaped cavity provide a simpler if not more plausible explanation for the observed X-ray emission.  The resulting entrained molecular gas would be expected to extend beyond the location of the X-ray source, as is observed for the compact central component.  To produce the observed X-ray luminosity and plasma temperature, \citet{bal03} find that a preshock gas density of 1--$10 \times 10^3 {\rm \ cm^{-3}}$ and shock velocities larger than $350 {\rm \ km \ s^{-1}}$ are required.  Such molecular hydrogen gas densities are required to excite the observed CO(2--1) emission from the compact central concentration.  The line-of-sight velocity of at least one of the two optical jets reach $\sim$$300 {\rm \ km \ s^{-1}}$, and if inclined by $\sim$$30\degr$ to the plane of the sky as inferred by \citet{lim06} would therefore reach true velocities of $\sim$$600 {\rm \ km \ s^{-1}}$.  Such high jet velocities can produce the required shock velocities.

Instead of tracing an outflow, could the compact central component constitute the inner region of the flattened molecular condensation around L1551~IRS5?  In such a case, the observed NE-SW velocity gradient would then correspond to infalling motion as seen in C$^{18}$O(1--0) \citep{mom98}.  The measured velocity gradient caused by the infalling motion at the same scale as the compact CO(2--1) central component ($\sim$300~AU) is at most $\sim$$0.13 {\rm \ km \ s^{-1} \ arcsec^{-1}}$ \citep[see Fig.~3 in][]{mom98}.  For comparison, the measured velocity gradient of  the compact central component is at least $\sim$$0.60 {\rm \ km \ s^{-1} \ arcsec^{-1}}$.  Furthermore, as shown by \citet{tak04}, the inner region of the condensation is dominated by rotation and not infall, and exhibits a velocity gradient in the NW-SE direction (along the major axis of the condensation) and not NE-SW direction.  Thus, the compact central component is unlikely to comprise the inner regions of the condensation around L1551~IRS5. 

\subsection{S-shaped Component}
The S-shaped component would not have been spatially resolved and hence recognized in previous single-dish CO observations.  Given that it exhibits a velocity pattern opposite to the compact central component, X-shaped component, and large-scale bipolar molecular outflow, at first sight it is unlikely to comprise material entrained by the collimated winds of the two main protostellar components.  Furthermore, this component exhibits the lowest (projected) maximum detectable velocity of all three components.  Below, we consider three possible explanations for this component: i) the near and far sides of the outflow cavity walls; ii) a precessing outflow; and iii) an outflow that winds in space due to the orbital motion of its driving source.

\subsubsection{Near and Far Sides of Outflow Cavity Walls}
Here, we consider the possibility that the S-shaped component is part of 
the same outflow cavity traced by the X-shaped component.  In the model described in $\S4.1$, the outflow cavity has a half-opening angle ($\sim$45\degr) that is larger than its inclination to the plane of the sky ($\sim$30\degr).  On the NE side of the outflow cavity where the observed limb-brightened walls are redshifted, the walls on the near side of this cavity is therefore blueshifted.  Conversely, on the SW side of the outflow cavity where the observed limb-brightened walls are blueshifted, the walls on the far side of this cavity is redshifted.  The S-shaped component may therefore comprise the near side of the outflow cavity on the NE side and the far side of the outflow cavity on the SW side.

In such a case, the model described in $\S4.1$ that fits the observed morphology and kinematics of the X-shaped component should also fit the observed morphology and kinematics of the S-shaped component.  Figure 9 shows the PV-diagram along the symmetry axis (position angle of 70$\degr$) of the X-shaped component.  Emission close to the center straddled by the horizontal short-dashed lines is dominated by the compact central component.  The solid line shows what our model described in $\S4.1$ (with $C = 0.18 {\rm \ km \ s^{-1} \ arcsec^{-1}}$ and $i = 30\degr$) predicts if the S-shaped component comprises the near side of the outflow cavity on the NE side and the far side of the outflow cavity on the SW side as described above.  Clearly the S-shaped component cannot be fit by such a model.  The PV-diagram of the S-shaped component in Figure~9 is better fit by the long-dashed line, which corresponds to an acceleration constant for a Hubble-like radial expansion of $C = 0.75 {\rm \ km \ s^{-1} \ arcsec^{-1}}$ for the same assumed inclination of $i = 30\degr$.  The long-dashed line in Figure~5 shows what this model predicts for the PV-diagram of the X-shaped component.  Clearly, the X-shaped component cannot be fit by such a model.

To simultaneously fit the PV-diagrams of both the X- and S-shaped components in a single model for the cone-shaped cavity, this cavity is required to have an inclination angle of $i$ = 16$\degr$ and acceleration constant for its Hubble-like radial expansion of $C = 0.36 {\rm \ km \ s^{-1} \ arcsec^{-1}}$.  With this set of parameters, the model fit to the X-shaped component is closely indicated by the solid line in Figure~5 and to the S-shaped component the dashed line in Figure~9.  The symmetry axis of such a cone-shaped cavity would then be tilted by $\sim$14\degr\ from a direction orthogonal to the circumstellar disks of the main northern and southern protostellar components \citep{lim06} as well as the surrounding envelope \citep{mom98}.  This model does not provide an obvious explanation for why the far wall on the NE side and near wall on the SW side are not detected.  More problematically, it does not provide a natural explanation for the S-shaped morphology of this component nor its oscillatory line-of-sight velocity as described next.

\subsubsection{A Precessing Outflow}\label{Precessing Outflow}
The morphology of the S-shaped component resembles a precessing outflow.  To assess whether its kinematics also resemble a precessing outflow, we adopt the following simple model.  We assume that the outflow is driven by a jet with a constant velocity $v_{jet}$ that precesses at a constant angular frequency with a period $T_{p}$.  Such a jet will describe a sinusoidal pattern in the sky with a deflection amplitude that increases with distance from the driving source in the form \citep[adapted from][]{eis96}:
\begin{equation}
\left(
\begin{array}{c}
x\\
y
\end{array}\right)
=\left(
\begin{array}{cc}
cos\psi&-sin\psi\\
sin\psi&cos\psi
\end{array}\right)
\left(
\begin{array}{c}
\alpha\cdot l\cdot \sin(2\pi l / \lambda+\phi_0)\\
l\cdot \cos i
\end{array}\right)  \ \ ,
\end{equation}
where $x$ and $y$ are the usual Cartesian coordinates, $\alpha$ the precession amplitude, $\lambda$ the precession length scale, $l$ the distance from the source, $\phi_0$ the initial phase at the source, $\psi$ the position angle of the outflow symmetry (precession) axis in the plane of the sky, and $i$ the inclination angle of the outflow symmetry axis to the plane of the sky.  The line-of-sight velocity, $V_{\rm LOS}$, can be derived from the equation:
\begin{equation}
V_{LOS}=V_{LSR}\pm V_{CO}[\cos\theta\sin i+\sin\theta\cos i\cos(2\pi l / \lambda+\phi_0)] \ \ ,
\end{equation}
where $V_{CO}$ is the measured CO(2-1) velocity, and $\theta$ the half-opening angle of the winding outflow.  Figure~10 shows a schematic picture of this model in which the jet precesses clockwise about its symmetry (precession) axis.  Note that smaller inclinations ($i$) require smaller precession amplitudes ($\alpha$) to produce a given line-of-sight velocity ($V_{\rm LOS}$) deflection in the sky.  

We start by considering a straight jet.  To fit the observed morphology and kinematics, we find that the inclination of the jet symmetry axis to the plane of the sky is restricted to the range $15\degr < i < 45\degr$. For illustration, we take $i \simeq 30\degr$, allowing us to directly compare the energetics of the S-shaped component with the X-shaped and compact central components (see $\S3.2.5$).  For this inclination, we find that the model parameters listed in Table~4 (first row) provide the closest fit to both the observed morphology and kinematics of the S-shaped component.  Within the allowed inclination range, other sets of model parameters can provide a more or less equally satisfactory fit to the observed morphology and kinematics.  

Figure~11 compares the model predictions (black line) in the case where $i \simeq 30\degr$ with the observed morphology (upper panels) and line-of-sight velocity (lower panel) of the S-shaped component.  Here, the jet that drives the S-shaped component precesses at an angle of $\theta \approx 22\degr$ to its symmetry (precession) axis.  The morphology of the S-shaped component is well reproduced in this model where the driving jet lies outside the conical outflow cavity traced by the X-shaped component along most of its length, thus enabling it to entrain the surrounding material.  The line-of-sight velocities plotted in Figure~11 (lower panel) are derived by fitting the measured spectra at regularly spaced intervals along the predicted path of the precessing outflow with simple gaussians.  Two sets of error bars are plotted for each point, the smaller (solid line) corresponding to the uncertainty in the measurement of the central velocity, and the larger (dashed line) the FWHM of the fitted gaussian.  As can be seen, the line-of-sight velocity also changes more or less sinusoidally in reasonable --- albeit not perfect --- agreement with the model.  Given that the entrained material may have had different initial velocities along the length of the outflow reflecting the rotation and contraction of the surrounding envelope, significant differences between the predicted and measured velocities may not be unexpected.    

The precession period, $T_p$, can be determined from the wavelength $\lambda$ and jet velocity $v_{jet}$ such that 

\begin{equation}
T_p \ ({\rm yrs}) = 4.76\times\frac{\lambda\times d ({\rm pc})}{v_{jet} ({\rm km \ s^{-1}})} \ \ ,
\end{equation}
\\
where $d$ is the distance to L1551~IRS5 (140~pc).  Assuming a typical jet velocity of $\sim$50--$150 {\rm \ km \ s^{-1}}$ (see below for reason why these values were adopted), the precession period is $\sim$130--400~yrs.

To see if we can produce an even better fit to the observed morphology and, especially, kinematics, of the S-shaped component, we relax the requirement that the jet driving this component be straight.  A bent jet (see inset in Fig.~10) --- in which the two opposing arms are not aligned --- have been seen or invoked to explain the bipolar molecular outflows from a number of protostellar systems.  The mechanisms that can lead to such bending have been explored by \citet{fen98}; in the present situation, wide-angle winds at the base of the main northern and/or southern protostellar components pushing on the jet of the third protostellar component provides yet another possibility.  Adopting once again $i \simeq30\degr$, we find that the model parameters listed in Table~4 (second row) provide a better fit to the observed kinematics.  Figure~12 plots the model predictions (black line) for the observed morphology (upper panels) and line-of-sight velocity (lower panel) of the S-shaped component, where the jet that drives this component precesses at an angle of $\sim$$10\degr$ to its symmetry axis.  As can be seen, the measured line-of-sight velocities along the precessing outflow are in relatively good agreement with model predictions for a bent jet where one arm lies at an angle of $163\degr$ to the other (for a straight jet, the two arms lie at an angle of $180\degr$).  For a jet velocity of $\sim$$100 {\rm \ km \ s^{-1}}$ as assumed above, the precession period corresponds to $\sim$135~yrs.

Could the S-shaped component, which as we have shown can be well fitted by a precessing outflow, be driven by the candidate third protostellar component in L1551~IRS5 discovered by \citet{lim06}?   The circumstellar disk of this third component is tilted with respect to the closely aligned circumstellar disks of the two main protostellar components, as well as their surrounding pseudodisk.  The outflow driven by a jet from the third protostellar component should therefore have a different inclination and/or position angle in the sky compared with the outflows driven by the northern and southern protostellar components.  Indeed, the S-shaped outflow is tilted in the opposite sense of the sky with respect to the compact central outflow, X-shaped outflow cavity, and large-scale bipolar molecular outflow, all of which are driven by the two main protostellar components.  In the model incorporating a straight jet, the jet currently emerges at its origin at a position angle of $78\degr$ in the plane of the sky.  By comparison, the major axis of the circumstellar disk of the third component is at a position angle of $118\degr \pm 8\degr$, and hence the jet is bent by $\sim$$50\degr$ from orthogonal to the disk.  In the model incorporating a bent jet, at or close to its origin the eastern (blueshifted) arm of the jet emerges at a position angle of $\sim$$65\degr$ and the western (redshifted) arm of the jet at a position angle of $\sim$$48\degr$ as sketched in the insert of Figure~10, within $\sim$$20\degr$ and $\sim$$37\degr$ respectively from orthogonal to the disk.

Tidal interactions between the third protostellar component and its more massive neighbor(s) could cause the disk of this component to precess, resulting in the precessing outflow seen.  \citet{bat00} have shown that such interactions between two protostellar components in a binary system should cause their circumstellar disks to precess at a frequency that is a small fraction (of order a few percent) of their orbital angular frequency.  Taking a mass for the main northern protostellar component of $0.45 \ M_\sun$ and the third protostellar component of $0.1 \ M_\sun$, and a physical separation between the two components equal to their projected separation of $\sim$13~AU \citep[see][]{lim06}, the orbital period of the third protostellar component about the northern protostellar component is $\sim$63~yrs.  For comparison, the estimated precession period of the S-shaped outflow is a few hundred years, assuming a jet velocity for the third protostellar component that is proportional to the square-root of its mass (i.e., proportional to the Keplerian velocity).  The results are compatible with the idea that the circumstellar disk of the third protostellar component is precessing at an angular frequency much smaller than the orbital angular frequency of this component about the main northern protostellar component.

\subsubsection{A Winding Outflow due to Orbital Motion}
As mentioned by \citet{fen98}, the orbital motion of a protostar about its binary companion could also cause its originally straight jet to be deflected backwards.  Because such a bent jet possesses mirror rather than point symmetry (as in the case of a precessing jet), it should drive a winding outflow with reflection symmetry \citep[e.g.,][]{mas02}.  We have therefore examined whether the morphology and kinematics of the S-shaped component can be fit by a winding mirror-symmetric outflow.  

For simplicity, we consider a circular orbit.  To produce the observed morphology, the orbital velocity, $v_{orb}$, of the protostar that drives the jet is required to satisfy the relation:
\begin{equation}
{v_{orb}\cos j \over v_{jet}\cos i} = \tan \theta \ \ ,
\end{equation}
where as before $v_{jet}$ is the jet velocity, $i$ the inclination to the plane of the sky at which the jet emerges from its driving source (i.e., tangent to the jet at its origin), $\theta$ the half-opening angle of the winding outflow (also equal to the angle at which the jet is bent from its original trajectory), and $j$ the orbital inclination to the plane of the sky.  The resulting line-of-sight velocity of such a jet will oscillate sinusoidally about the jet velocity, with the oscillation caused by the orbital motion of the protostar driving the jet.  As can be seen from Figure~12, the molecular outflow oscillates with a half amplitude of $\sim$$0.5 {\rm \ km \ s^{-1}}$ about a line-of-sight velocity of $\sim$$1.5 {\rm \ km \ s^{-1}}$; i.e., ratio between these two values of $\sim$0.3.  To produce the observed kinematics, the orbital velocity of the protostar is therefore required to also satisfy the relation:
\begin{equation}
{v_{orb}\sin j \over v_{jet}\sin i} \approx 0.3 \ \ .
\end{equation}
We assume throughout that the orbital diameter of the protostar is much smaller than the observable width of the winding outflow (i.e., that the jet emerges from a stationary point), an assumption that we shall return to later.

We found that a mirror-symmetric outflow model can provide a reasonable fit to the observed morphology and kinematics of the S-shaped component over only a relatively narrow range in values of $14\degr \lesssim \theta \lesssim 19\degr$ and $40\degr \lesssim i \lesssim 50\degr$.  For illustration, we take median values of $\theta \approx 17\degr$ and $i \approx 45\degr$, in which case Equations~(5) and (6) can be solved to give $j \approx 45\degr$.  To fit the observed morphology and kinematics, we also require $\lambda = 17\arcsec$.  The model fit for these values is shown in Figure~13.  It is superior, especially in reproducing the observed kinematics, to the model fit for a straight precessing jet, but inferior to that for a bent precessing jet. 

The orbital period of the protostar is equal to the period of the winding outflow, which has the same dependence on the model parameters as the precession period given by Equation~(4).  Substituting into this equation $v_{jet} \approx v_{orb} / 0.3$ from Equation~(6) given the model parameters used here, we find $T_{orb} (yrs)=0.3 \times 4.76\times \lambda \times d(pc) / v_{orb}$.  The product $v_{orb} \times T_{orb}$ is simply the circumference of the orbit, from which we derive an orbital diameter of $\sim$236~AU ($\sim$1\farcs7).  Thus, the orbital diameter is indeed much smaller than the observable width of the winding outflow, satisfying an underlying assumption of our model.  In this model, the minimum projected orbital radius is $\sim$83~AU ($\sim$0\farcs6), many times larger than the observed projected separation of the 3rd component from the main northern protostellar component of $\sim$13~AU ($\sim$0\farcs1).  Indeed, the minimum projected orbital radius is a factor of $\sim$2 larger than the observed projected separation of the 3rd component from the main southern protostellar component.  Thus, to fit all the observational constraints, elliptical orbits with high eccentricities would have to be considered (or a fourth yet unseen protostellar component invoked).

For the model parameters used here, the orbital period can also be expressed as
\begin{equation}
T_{orb} \ ({\rm yrs}) \approx \frac{1290 (\rm yrs)}{[M_N+M_{3rd} (\rm M_{\sun})]^{1/2} } \ \ ,
\end{equation}
where M$_N$ and M$_{3rd}$ are the masses of the main northern and 3rd protostellar components respectively.  Taking M$_N \approx 0.45 {\rm \ M_{\sun}}$ and M$_{3rd} \approx 0.1 {\rm \ M_{\sun}}$ \citep{lim06}, we find $T_{orb}\approx 1700 {\rm \ yrs}$ and hence $v_{orb}\approx 2.1 {\rm \ km \ s^{-1}}$ and $v_{jet} \approx 7 {\rm \ km \ s^{-1}}$.  The jet velocity in this model is unreasonably small, a serious flaw in the picture where the winding outflow is due to the orbital motion of the driving source.

\section{Summary}
L1551~IRS5 exhibits the first recognized bipolar molecular outflow from a protostellar system.  Single-dish maps show that this outflow can be traced in both CO(1--0) and CO(2--1) to 1.4~pc on the sky along the NE (redshifted lobe) and SW (blueshifted lobe) sides of the system along a position angle of $\sim$$50\degr$.  This outflow is presumably driven by the two main protostellar components in L1551~IRS5, each of which exhibits an ionized jet that is closely aligned with the major axis of the larger scale molecular outflow.

Here, we imaged the molecular outflow within $\sim$4000~AU of L1551~IRS in CO(2--1) with the SubMillimeter Array (SMA)  at an angular resolution of $\sim$4\arcsec\ (linear resolution of $\sim$560~AU).  Our observations revealed three distinct spatial-kinematic components:

\begin{itemize}

\item[1.]  An X-shaped structure centered on L1551~IRS with arms that can be traced to $\sim$20\arcsec--30\arcsec\ ($\sim$2800-4200~AU) from the system.  It has a symmetry axis along the NE (redshifted side) to SW (blueshifted side) direction at a position angle of $\sim$$70\degr$.  This component is therefore closely aligned with, and exhibits the same velocity pattern as, the large-scale bipolar molecular outflow.

\item[2.]  A previously unknown S-shaped structure centered on L1551~IRS5 that can be traced to $\sim$15\arcsec\ ($\sim$2100~AU) from the system.  It has a symmetry axis along the NE (blueshifted side) to SW (redshifted side) direction at a position angle of $\sim$$55\degr$.  Although also closely aligned with the major axis of the large-scale bipolar molecular outflow, this component exhibits a velocity pattern in the opposite sense.  There is no known counterpart to this outflow on large ($\gtrsim 0.1$~pc) scales.  This component has the smallest linewidth of all three components.

\item[3.]  A previously unknown compact component centered on L1551~IRS5 that can be traced to $\sim$1.4\arcsec\ ($\sim$200~AU) from the system.  It has a major axis along the NE (redshifted side) to SW (blueshifted side) direction at a position angle of $\sim$$65\degr$.  This component is therefore closely aligned with, and exhibits the same velocity pattern as, the large-scale bipolar molecular outflow.  It has the largest linewidth of all three components.

\end{itemize}

We interpreted our results in the following manner:

\begin{itemize}

\item[1.]  The X-shaped component comprises the limb-brightened walls of a cone-shaped outflow cavity.  The molecular gas originally in this cavity was likely entrained by the high-velocity collimated winds of the two main protostellar components, and subsequently excavated by the low-velocity wide-angle winds of the same components.  The opening angle of this cavity is $\sim$90$\degr$, comparable to that seen in other Class~I protostellar systems.

\item[2.]   The S-shaped component is unlikely to be the front and back walls of the same outflow cavity whose limb-brightened walls are seen as the X-shaped component.  Instead, the S-shaped component is most likely a precessing outflow whose symmetry axis is inclined in the opposite sense to the plane of the sky compared with the X-shaped and compact central components, as well as the large-scale bipolar molecular outflow.  This precessing outflow may driven by the candidate third protostellar component in L1551~IRS5 discovered by \citet{lim06}, who found that the circumstellar disk of this component is misaligned relative to the two main protostellar components.  Gravitational interactions between the third protostellar component and its likely more massive northern (and perhaps also southern) neighbor(s) may be causing the circumstellar disk and hence jet of this component to precess. 

\item[3.]  The compact central component likely comprises material newly 
entrained by one or both jets from the two main protostellar components.  The relatively short dynamical age for this component of $\sim$200~yrs may suggest a recent enhancement in the strength of one or both ionized jets responsible for driving this outflow.

\end{itemize}

Previous single-dish observations suggest multiple ejection episodes in the large-scale bipolar molecular outflow of L1551~IRS5 \citep{bac94}.  Our study reveals multiple bipolar molecular outflow components near the base of this large-scale outflow.  It adds additional evidence for a third protostellar component in L1551~IRS5 that may be related to that discovered by \citet{lim06}.

\acknowledgments
We thank Chin-Fei Lee for his advice in constructing the model for the X-shaped cavity, and N. Ohashi and M. Saito for fruitful discussions.  We thank the anonymous referee for suggesting we consider whether an outflow that winds due to orbital motion can explain the S-shaped component, and the editor (E. Fiegelsen) for bringing to our attention the possibility that the X-ray source seen at the base of the optical jets could be related to the compact central concentration.  J. Lim and S. Takakuwa acknowledge grants from the National Science Council of Taiwan (NSC 96-2112-M-001-020-MY2 and NSC 97-2112-M-001-003-MY2 respectively) in support of this work.  The grant to J. Lim also provides a Masters student stipend for P.-F. Wu.

{\it Facilities:} \facility{SMA}.

\clearpage
\begin{table}
\begin{center}
\caption{Observational Parameters}
\begin{tabular}{ll}
\hline\hline
PARAMETER & VALUE\\
\hline
Right Ascension (J2000) & 04$^{\rm h}$ 31$^{\rm m}$ 34$\fs$14\\
Declination (J2000) & 18$\arcdeg$ 08$\arcmin$ 05$\farcs$1\\
Primary beam (FWHM) & $\sim$57\arcsec\\
Synthesized beam (FWHM) & 4$\farcs$66 $\times$ 2$\farcs$30 (P.A.~$= -66.7\arcdeg$)\\
Central frequencies & 231.3~GHz, 241.3~GHz\\
Frequency (velocity) resolution & 203.125 kHz ($\sim$0.264 km s$^{-1}$)\\
Gain calibrators & 0423-013 (9.4 Jy), 3C~120 (3.3 Jy)\\
Absolute flux and passband calibrator & Uranus\\
rms noise level (continuum) & 0.014 Jy beam$^{-1}$\\
rms noise level (line) & 0.35 Jy beam$^{-1}$ at 203.125~kHz resolution\\
\hline
\end{tabular}
\end{center}
\end{table}

\clearpage
\begin{table}
\begin{center}
\begin{threeparttable}
\caption{Continuum Flux Density of L1551~IRS5}
\begin{tabular}{rcc}
\hline\hline
Wavelength & Flux Density (mJy) & References\tnote{a}\\
\hline
18 cm & 1.3 $\pm$ 0.4 & 1\\
3.6 cm & 1.48 $\pm$ 0.04 & 1\\
2.0 cm & 2.1 $\pm$ 0.2 & 1\\
1.3 cm & 3.5 $\pm$ 0.4 & 1\\
7 mm & 12.2 $\pm$ 1.0 & 1\\
7 mm & 10.1 $\pm$ 0.7 & 2\\ 
2.7 mm & 171 $\pm$ 19 & 3\\
2.7 mm & 162 $\pm$ 6 & 4\\
1.3 mm & 1200 & This work\\
850 $\mu$m & 7230 & 4\\
450 $\mu$m & 45700 & 4\\ 
\hline
\end{tabular}
\begin{tablenotes}
\footnotesize
\item[a]  (1) \citet{rod98}; (2) \citet{lim06}; (3) \citet{mom98}; (4) \citet{mo06}
\end{tablenotes}
\end{threeparttable} 
\end{center}
\end{table}

\clearpage
\begin{table}
\scriptsize
\begin{center}
\caption{Properties of the Three Outflow Components}
\begin{tabular}{ccccccc}
\hline\hline
      &      & Mass ($M$) & Momentum ($P$) & Energy ($E$) & Mass-loss rate ($\dot{M}$) & Dynamical Age \\
      &      & ($10^{-3} {\rm \ M_{\sun}}$) & ($10^{-3} {\rm \ M_{\sun} \ km \ s^{-1}}$) & (10$^{40}$~erg)
& ($10^{-7}$ $\rm {M_{\sun} \ yr^{-1}}$) & (yrs) \\
\hline
$X$-shaped & Red lobe & 0.48 & 2.28 & 12.12 & 2.60 & 1.3$\times10^4$ \\
               & Blue lobe & 0.77 & 4.07 & 25.25 & 3.37 & 1.2$\times10^4$ \\
$S$-shaped & Red lobe & 0.09 & 0.34 & 1.45 & 0.42 & 6.3$\times10^3$ \\
               & Blue lobe & 0.29 & 0.94 & 3.44 & 1.80 & 7.4$\times10^3$ \\
Central & Red lobe & 0.29 & 0.95 & 3.56 & 10.98 & 2.7$\times10^2$ \\
                  & Blue lobe & 0.22 & 1.33 & 9.66 & 11.00 & 2.0$\times10^2$ \\
\hline
\end{tabular}
\end{center}
\end{table}
\clearpage

\begin{table}
\begin{center}
\caption{Model parameters for precessing jet}
\begin{tabular}{lcccccc}
\hline\hline
     & $\psi$ & $\alpha$ & $\theta$  & $\lambda$ & $\phi_0$ (red, blue)& $i$ \\
\hline
Straight Jet & 57$\degr$ & 0.40 & 22$\degr$ & 40 & -90$\degr$, -90$\degr$ & 30$\degr$ \\
Bent Jet & 57$\degr$ & 0.18 & 10$\degr$ & 20 & 110$\degr$, -135$\degr$ & 30$\degr$ \\
\hline
\end{tabular}
\end{center}
\end{table}

\clearpage

\begin{figure}
\plotone{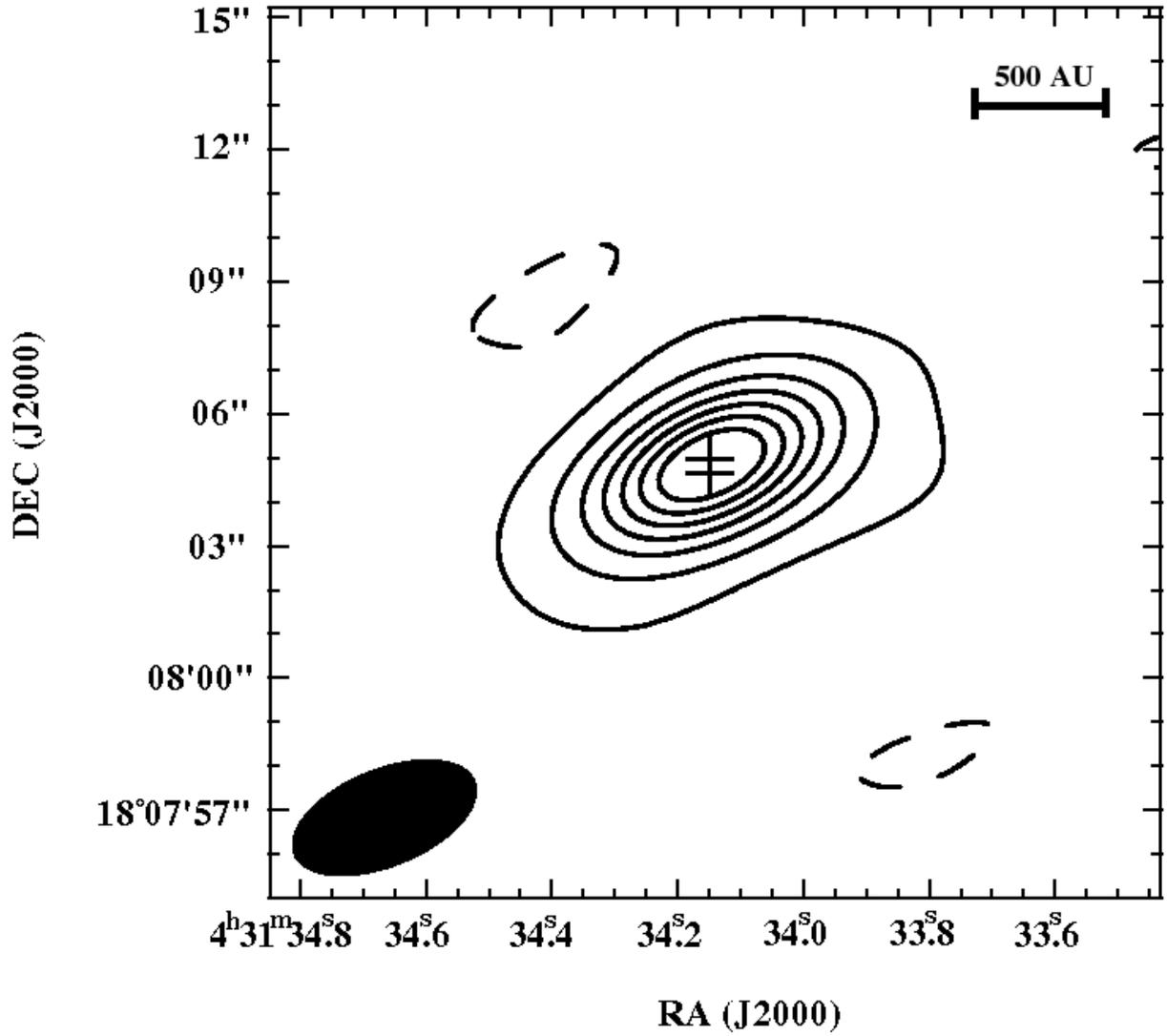}
\caption{Continuum map of L1551 IRS5 at 1.3 mm.  Contour levels are plotted at $-2$, 2, 10, 20, 30, 40, 50, 60$\sigma$, where the rms uncertainty $\sigma = 0.014 {\rm \ Jy \ beam^{-1}}$.  Crosses indicates the position 
of the two main protostellar components at 7~mm as measured by \citet{lim06}. 
The synthesized beam is shown as a filled ellipse at the bottom left corner.}
\end{figure}

\begin{figure}
\plotone{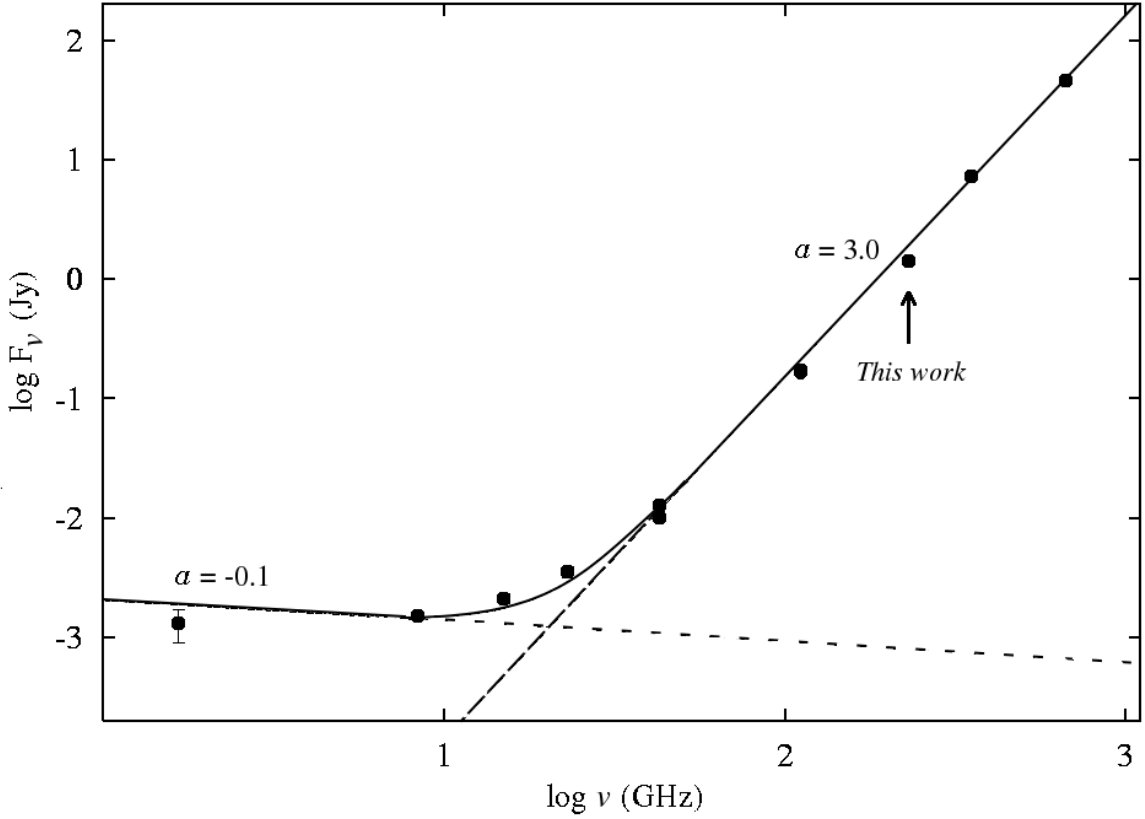}
\caption{
SED of the continuum emission from L1551~IRS5.  The data points and their corresponding $\pm 1\sigma$ measurement uncertainties (where large enough to be seen) are tabulated in Table~2.  The solid curve shows our best chi-squared fit assuming two power-law components with each having $F_\nu \propto \nu^{\alpha}$.  One component has $\alpha \approx -0.1$ as indicated by the short-dashed line and dominates at low frequencies, and the other $\alpha \approx 3.0$ as indicated by the long-dashed line and dominates at high frequencies (see text)}.
\end{figure}

\begin{figure}
\includegraphics[scale=0.7]{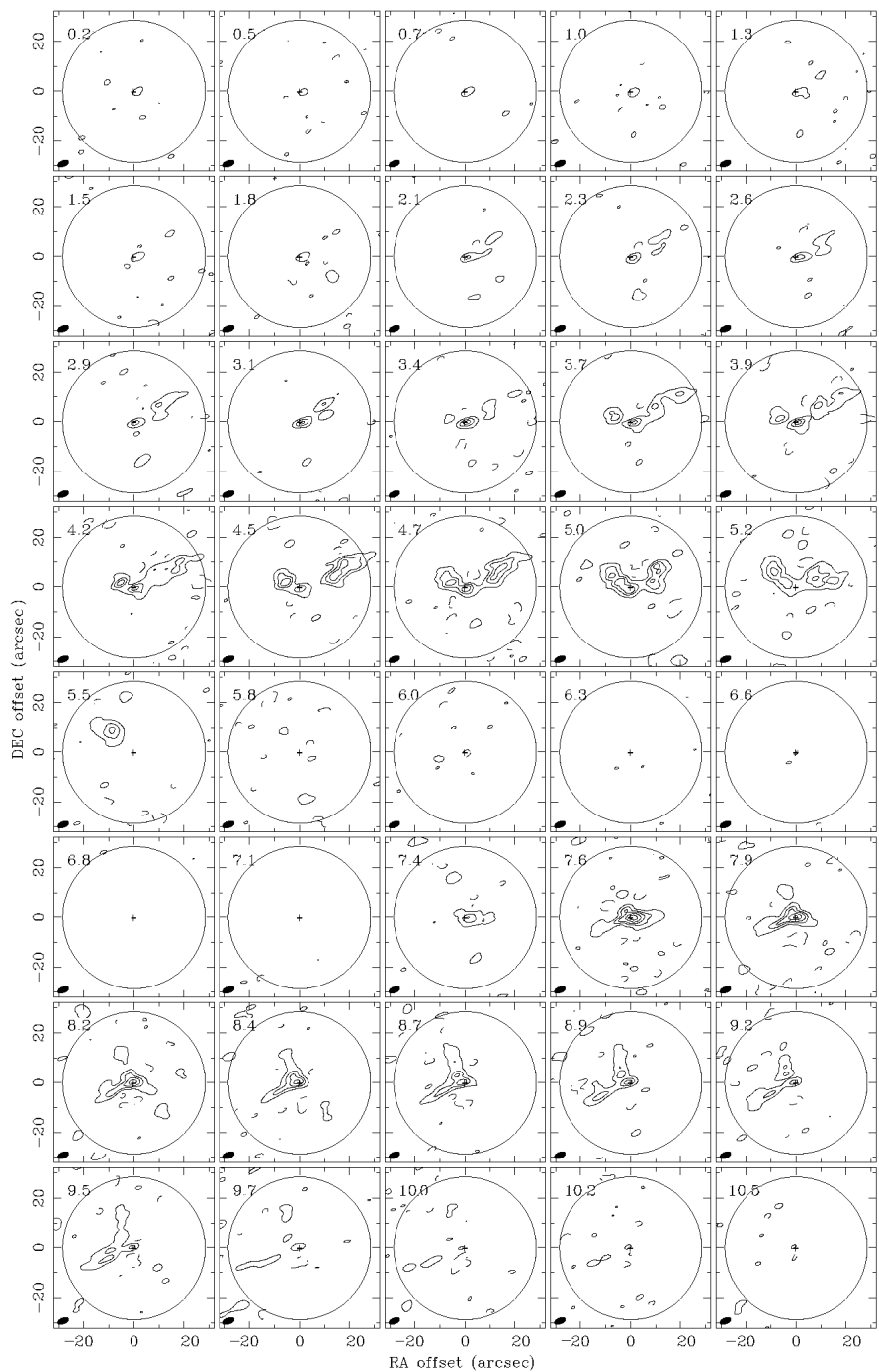}
\caption{CO(2--1) channel maps of L1551~IRS5. Contours levels are plotted at $-3$, 3, 6, 9, 15, 25, 35 and 45$\sigma$, where $\sigma = 0.35 {\rm \ Jy \ beam^{-1}}$.  The central $V_{\rm LSR}$ of each channel is shown at the top left corner of each panel.  Crosses indicate the positions to the two main protostellar components as in Figure~1.  The field of view of the SMA at FWHM is shown by the large circle, and the synthesized beam by the filled ellipse at the bottom left corner.}
\end{figure}

\begin{figure}
\plotone{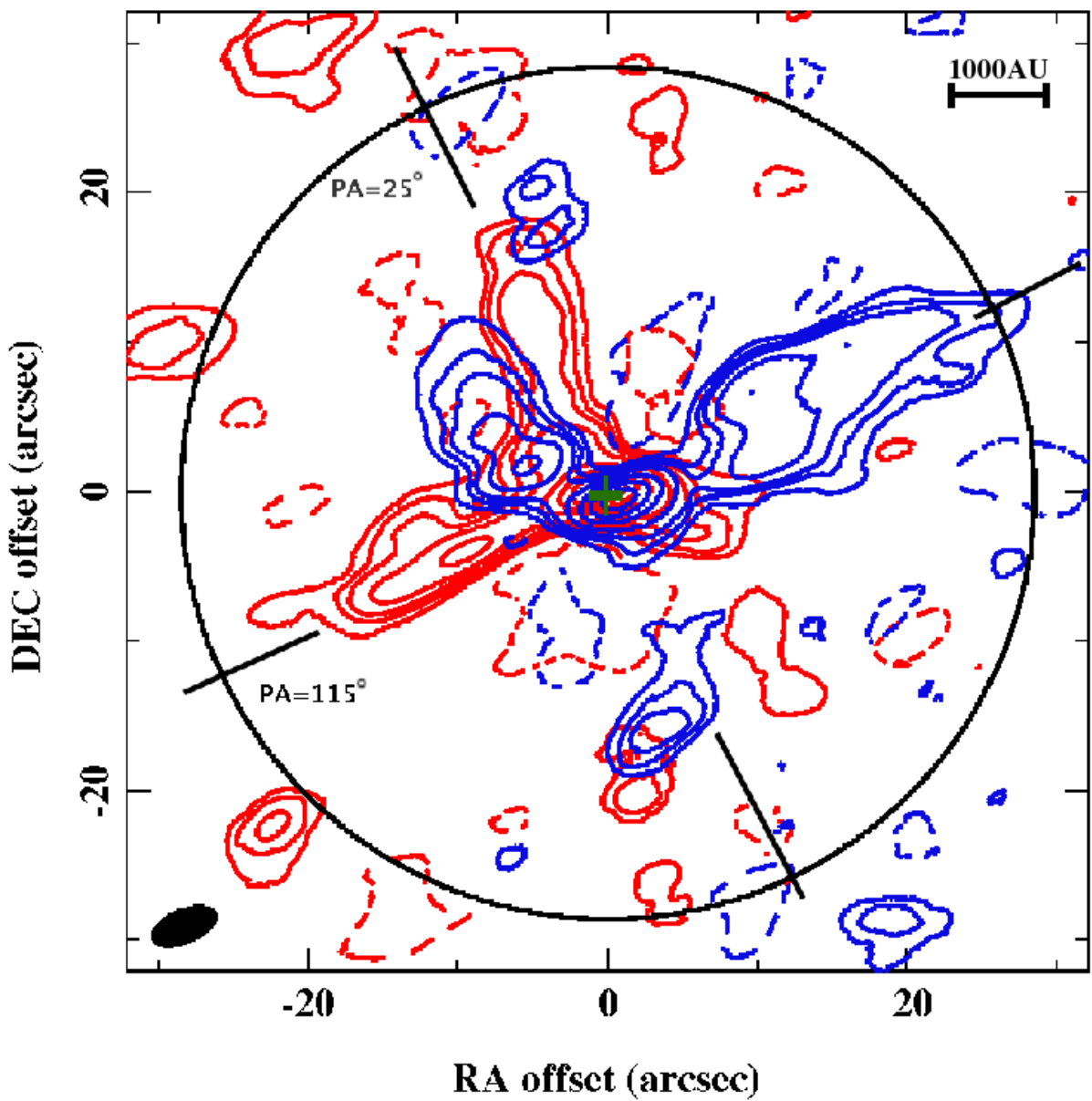}
\caption{Integrated CO(2--1) intensity map with blue contours showing emission at blueshifted velocities spanning V$_{\rm LSR} = 0.23$--$5.5 {\rm \ km \ s^{-1}}$, and red contours emission at redshifted velocities spanning V$_{\rm LSR} = 7.4$--$10.5 {\rm \ km \ s^{-1}}$.  For the blueshifted emission, contour levels are plotted at $-1.3$, 1.3, 2.6, 3.9, 6.4, 10.7, 15.0, and 19.3 ${\rm \ Jy \ km \ s^{-1} \ beam^{-1}}$, while for the redshifted emission, contour levels are at $-1.0$, 1.0, 2.0, 3.0, 5.1, 8.5, 11.9 and 15.3 ${\rm \ Jy \ km \ s^{-1} \ beam^{-1}}$.  The cross indicates the location of L1551~IRS5 protostellar system.  The field of view of the SMA at FWHM is shown by the large circle, and the synthesized beam by the filled ellipse at the lower left corner.  Diagonal lines indicate the orientation of cuts in the position-velocity diagrams shown in Figure~5.}
\end{figure}

\begin{figure}
\plotone{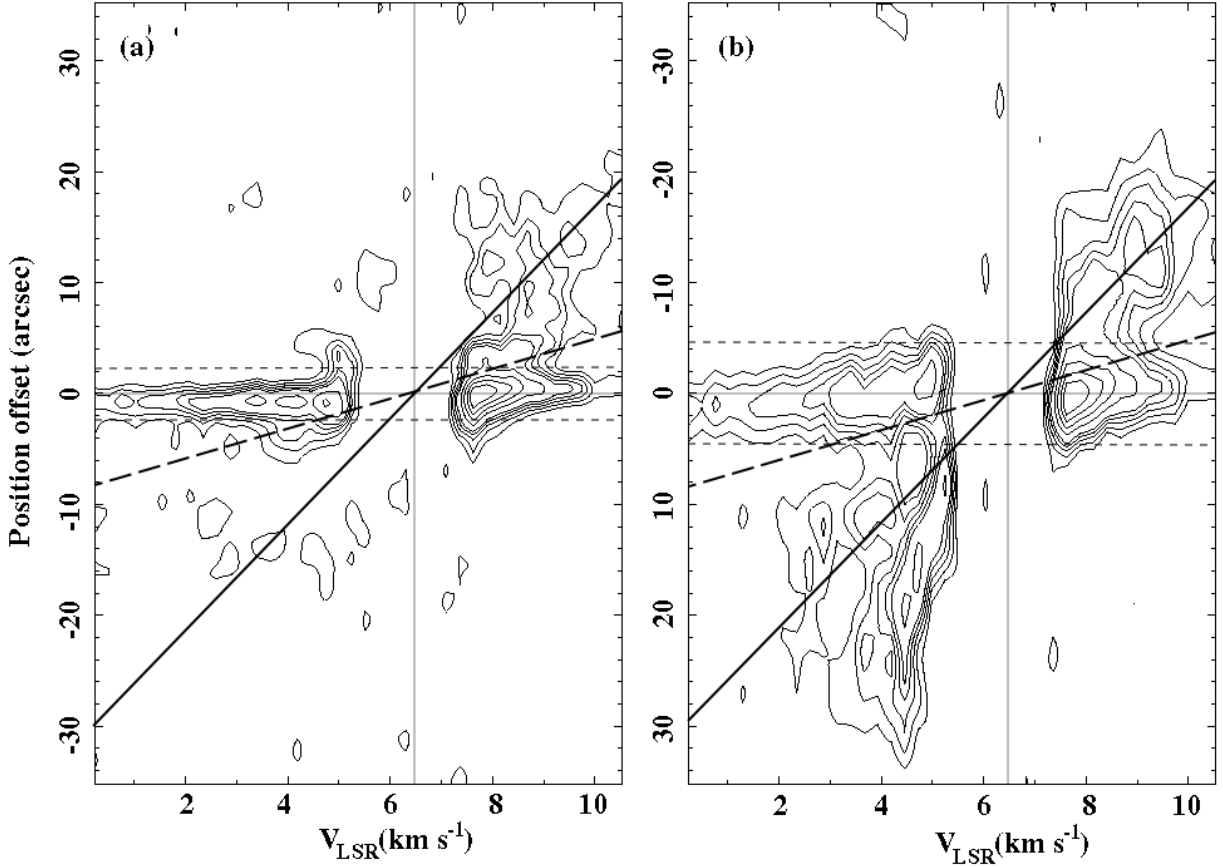}
\caption{PV-diagrams of the CO(2--1) emission along position angles (a) 25\degr\ and (b) 115\degr\ passing through the arms of the X-shaped component and central protostellar system (see Fig.~4).  Contour levels are plotted at 0.7, 1.4, 2.1, 2.8, 3.5, 5.5, 7.5, 9.5, and 11.5 Jy beam$^{-1}$.  In each panel, the pair of horizontal short-dashed lines straddling the horizontal solid line (which indicates the position of the protostellar system) are separated by twice the FWHM of the synthesized beam at that particular position angle.  Emission within this region is dominated by the compact central component (see text).  The vertical line indicates the systemic velocity that we inferred for L1551~IRS5.  The dark solid lines are the predictions of our model described in $\S4.1$.  The long-dashed lines are the predictions of a model described in $\S4.3.1$.  Note that panel (b) has been inverted vertically to show that three of the four arms (except the blueshifted NW arm) share the same velocity pattern (see text).}
\end{figure}

\begin{figure}
\plotone{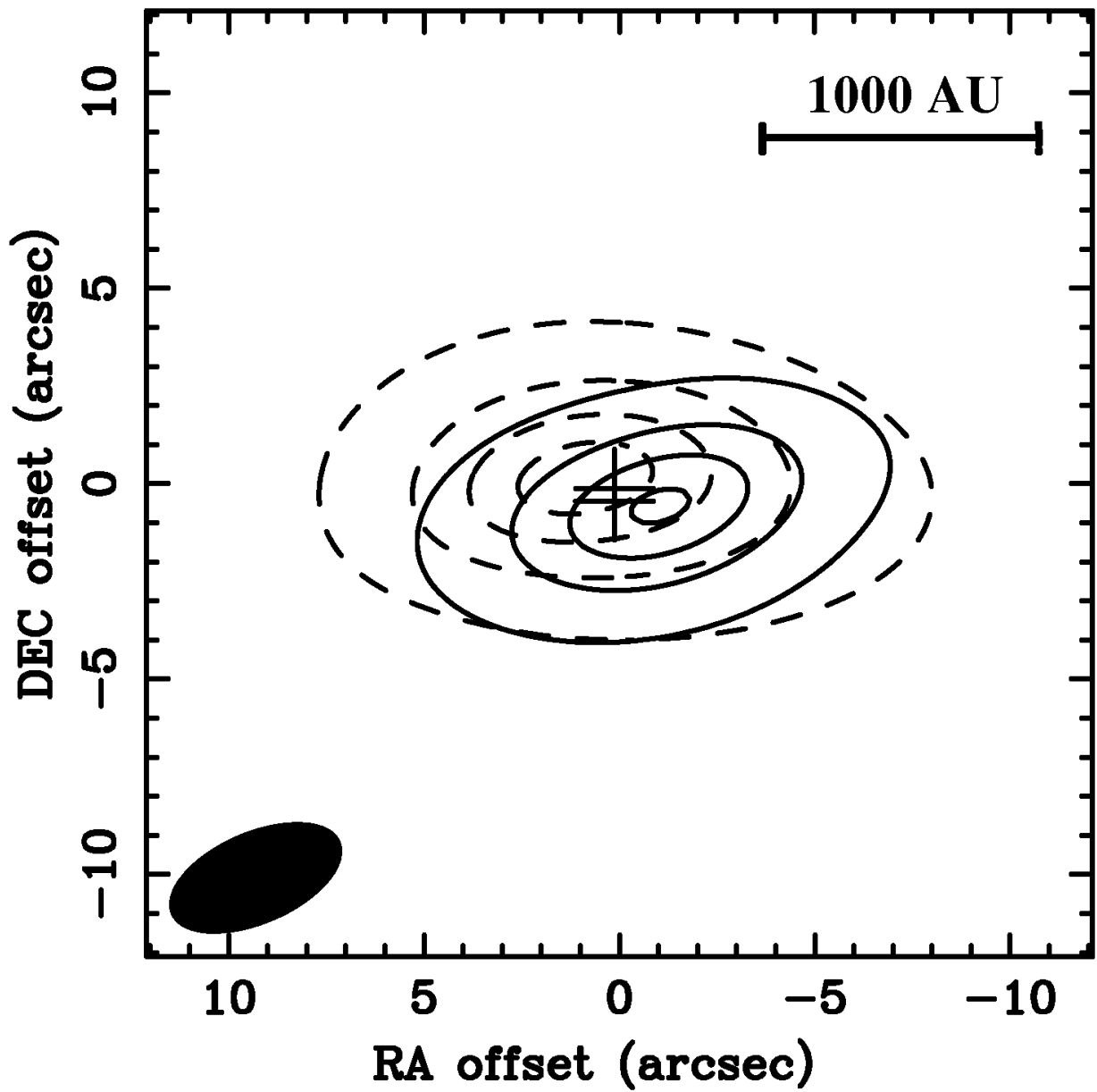}
\caption{Integrated CO(2-1) intensity map constructed from gaussian fits to the compact central component in the channel maps (see text).  Blue contours shows emission at blueshifted velocities spanning 0.23--$5.5 {\rm \ km \ s^{-1}}$, and red contours emission at redshifted velocities spanning 7.4--$10.5 {\rm \ km \ s^{-1}}$. Contours levels are plotted at $-3$, 3, 9, 15, 25, 35, and 45$\sigma$, where $\sigma = 0.43  {\rm \ Jy \ km \ s^{-1} \ beam^{-1}}$ for the blueshifted and $\sigma = 0.34 {\rm \ Jy \ km \ s^{-1} \ beam^{-1}}$ for the redshifted emission.  The synthesized beam is shown as a filled ellipse at the bottom left corner.}
\end{figure}

\begin{figure}
\includegraphics[scale=0.8]{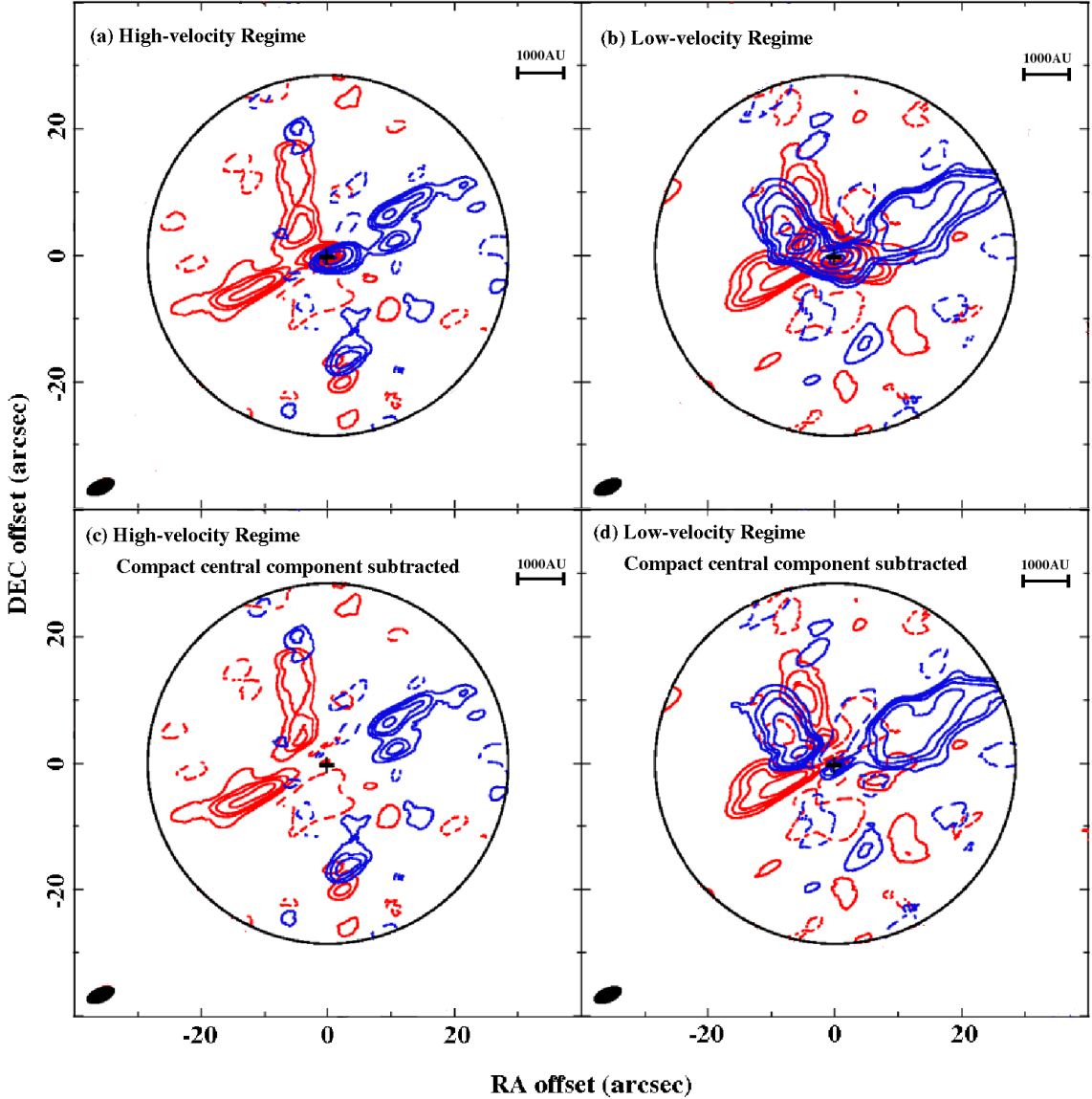}
\caption{Integrated CO(2--1) map with the compact central component included (upper row) and with it subtracted (lower row) separated into two velocity regimes (left and right columns).  In the higher velocity regime (left column), the blue contours span velocities 0.23--$0.34 {\rm \ km \ s^{-1}}$ and the red contours 8.9--$11 {\rm \ km \ s^{-1}}$.  In the lower velocity regime (right column), the blue contours span velocities 3.7--$5.5 {\rm \ km \ s^{-1}}$ and the red contours 7.4--$8.7 {\rm \ km \ s^{-1}}$.  In the left column, blue contours levels are plotted at $-1.0$, 1.0, 2.0, 3.0, 5.1, 8.5, 11.9 and 15.3 ${\rm Jy \ km \ s^{-1} \ beam^{-1}}$ and red contours levels are plotted at $-0.8$, 0.8, 1.5, 2.3, 3.8, 6.3, 8.8 and 11.3 ${\rm Jy \ km \ s^{-1} \ beam^{-1}}$. In the right column, blue contours levels are plotted at $-0.8$, 0.8, 1.6, 2.4, 3.9, 6.5, 9.1 and 11.7 ${\rm Jy \ km \ s^{-1} \ beam^{-1}}$ and red contours levels are plotted at $-0.7$, 0.7, 1.4, 2.1, 3.5, 5.8, 8.1 and 10.4 ${\rm Jy \ km \ s^{-1} \ beam^{-1}}$.}
\end{figure}

\begin{figure}
\plotone{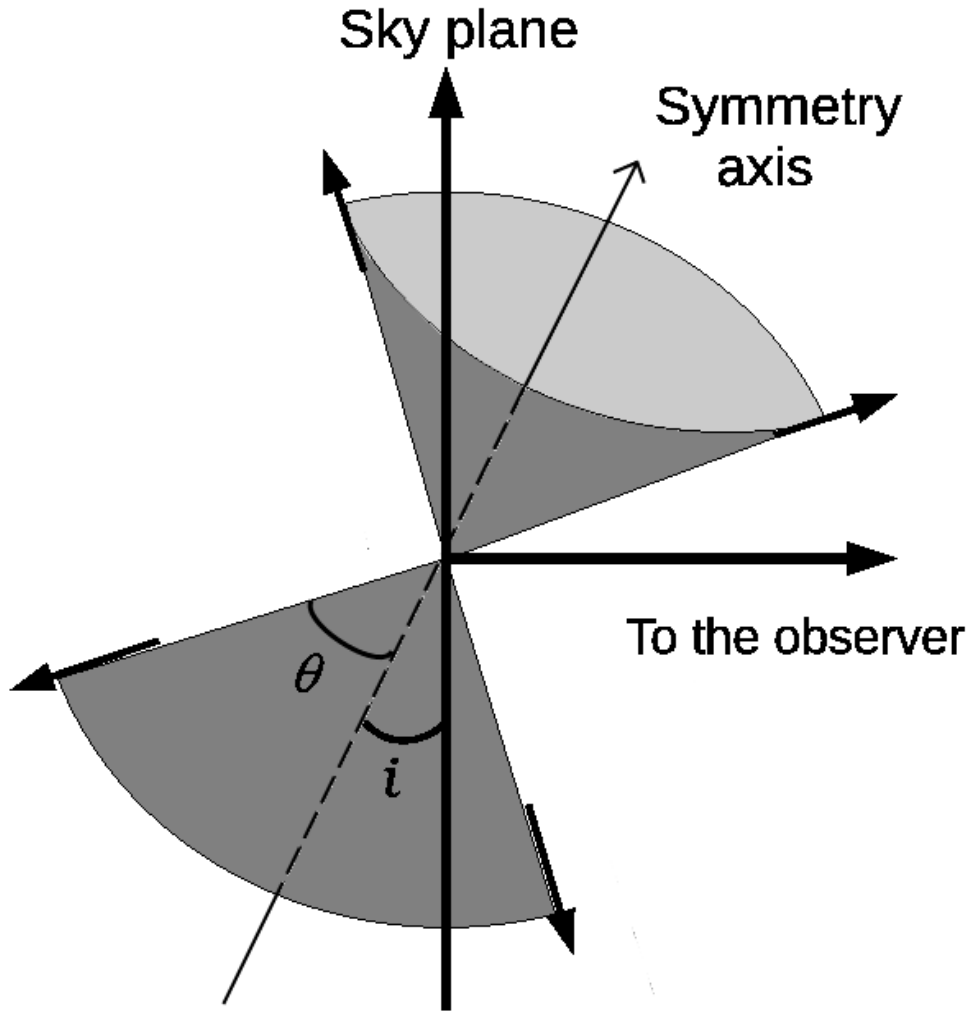}
\caption{A schematic picture of a cone-shaped outflow cavity that constitutes our model for the X-shaped component.  The cone has a half-opening angle of $\theta$, and is inclined by an angle $i$ to the plane of the sky.  The arrows indicate the direction of gas motion, which are along the cavity walls.}
\end{figure}

\begin{figure}
\includegraphics[scale=0.8]{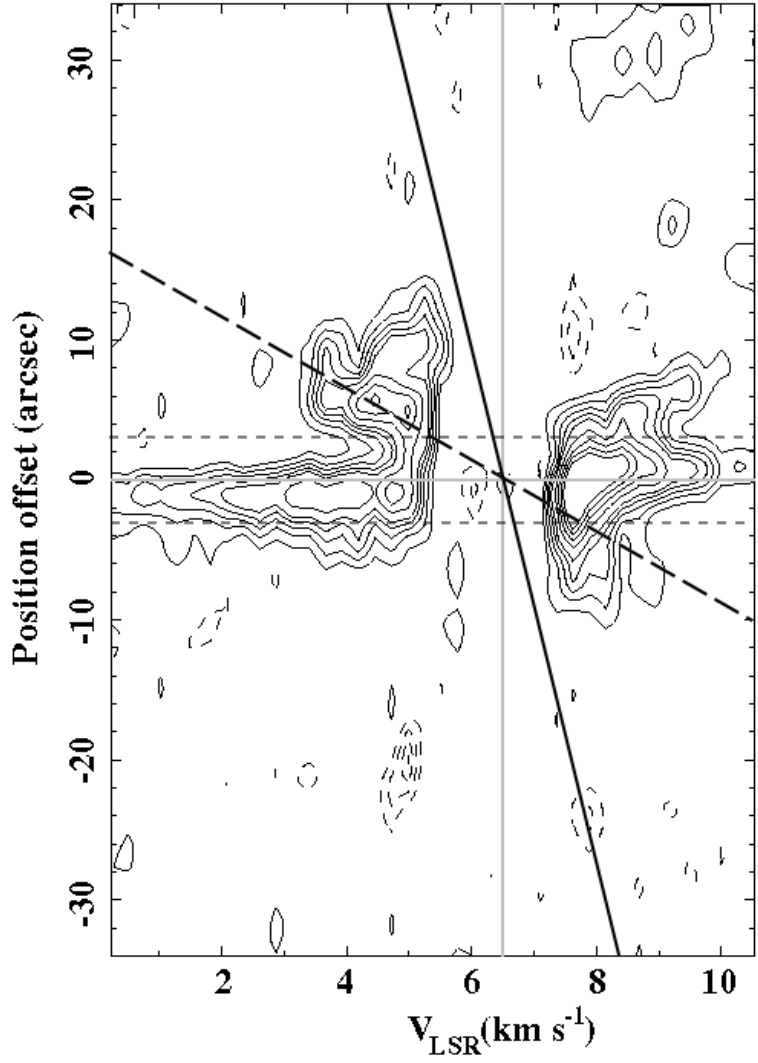}
\caption{PV-diagram of the CO(2--1) emission along the symmetry axis of the S-shaped component along a position angle of $= 70\degr$.  Contour levels and the different lines are the same as in Figure~5.}
\end{figure}

\begin{figure}
\begin{center}
\plotone{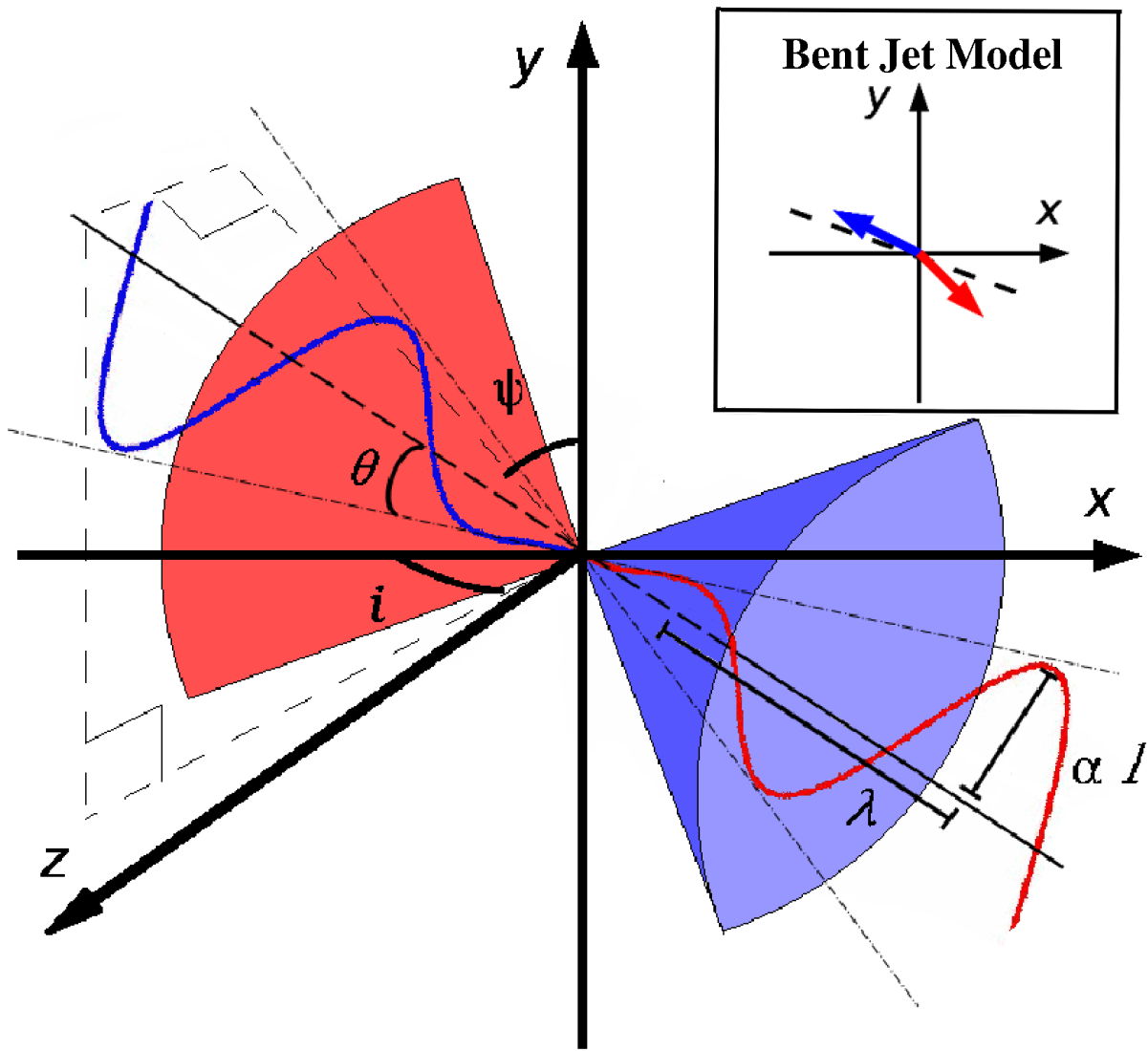}
\end{center}
\caption{A schematic picture of a precessing jet, represented by the blue/red sinusoidal curve, that constitutes our model for the S-shaped component.  The symmetry axis of the precessing jet is indicated by the thick dashed line, which forms an angle $i$ to the plane of the sky and lies at a position angle $\psi$ in the plane of the sky.  The amplitude of the precessing jet, $\alpha$$l$, increases with the distance $l$ away from the driving source.  The jet precesses on a cone with a half-opening angle of $\theta$, which is represented by two dash-dot lines.  The red and blue cones represent the outflow cavity traced by the X-shaped component (see Fig.~8), which has the opposite velocity pattern to the S-shaped component.  The blue/red arrows in the upper-right insert shows the directions in which a bent jet emerges from its driving source in the model discussed in $\S$4.3.2.  The black dashed line is orthogonal to the circumstellar disk of the 3rd protostellar component.}
\end{figure}

\begin{figure}
\begin{center}
\includegraphics[scale=0.8]{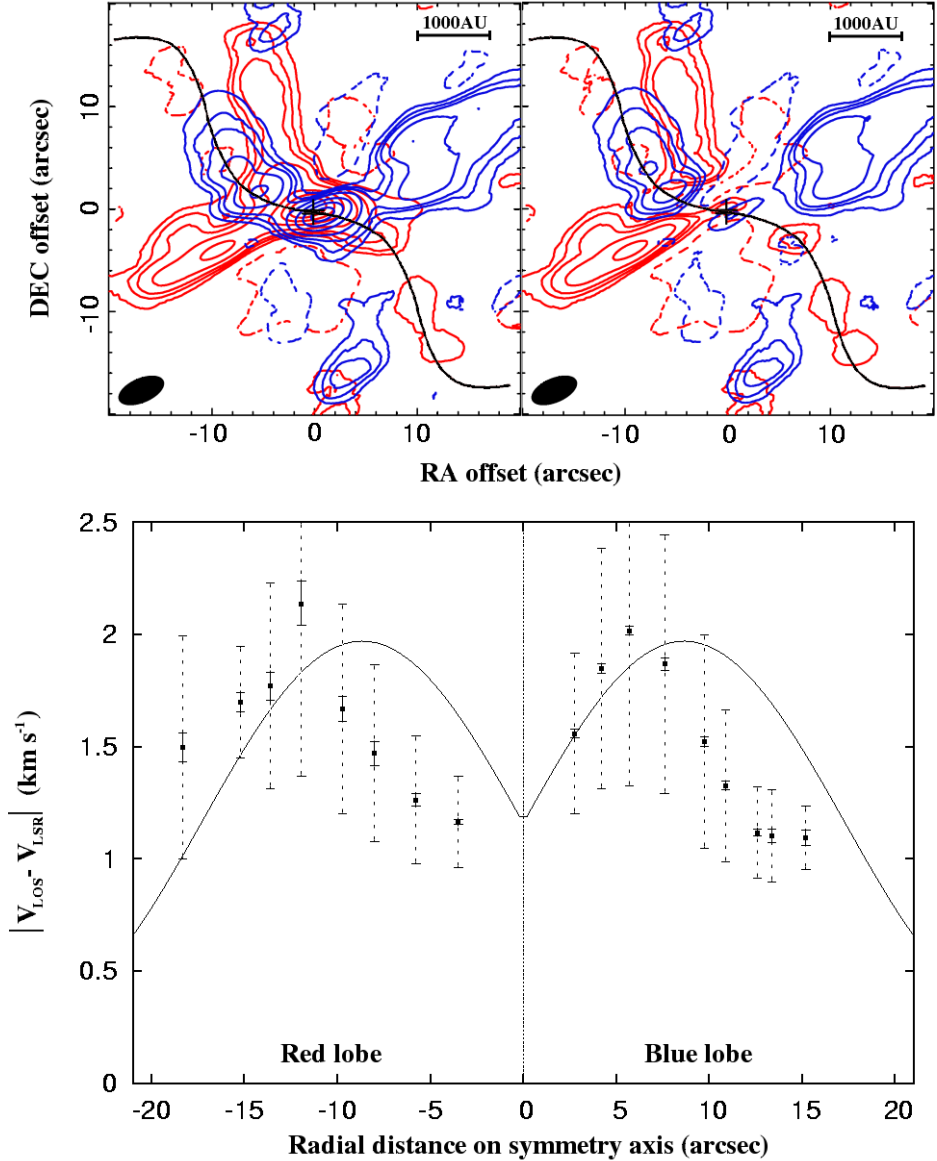}
\end{center}
\caption{Predicted morphology (upper panels) and line-of-sight velocity along the path (lower panel) of a precessing outflow, driven by a straight jet, with model parameters listed in the first row of Table~4.  The black curves in the upper panels show the path of the jet superposed on the integrated CO(2--1) intensity map of Figure~4 with the central compact component included (left panel) and with it subtracted out (right panel).  The black curves in the lower panels show the derived line-of-sight velocity of the outflow along the path of the jet superposed on the central CO(2-1) velocities derived from gaussian fits to the spectra at regularly spaced intervals along this path (see text).  The smaller solid error bars correspond to measurement uncertainties in the central velocities, and the larger large dashed error bars the FWHM of the fitted gaussians.}

\end{figure}

\begin{figure}
\begin{center}
\includegraphics[scale=0.8]{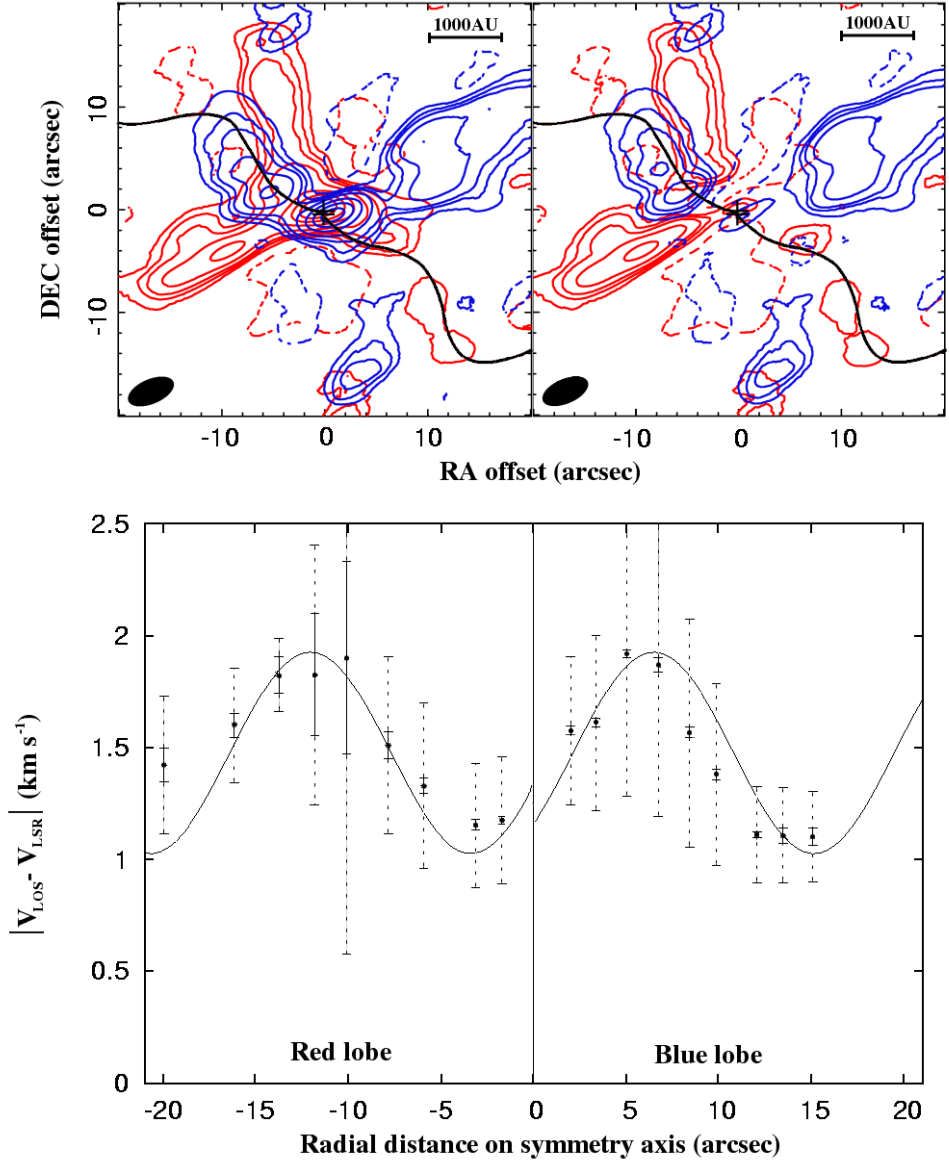}
\end{center}
\caption{Same as for Figure~11, but in this case for a precessing outflow driven by a bent jet with the model parameters listed in the second row of Table~4.  This model better reproduces the observed line-of-sight velocities as shown in the lower panel.}
\end{figure}

\begin{figure}
\begin{center}
\includegraphics[scale=0.8]{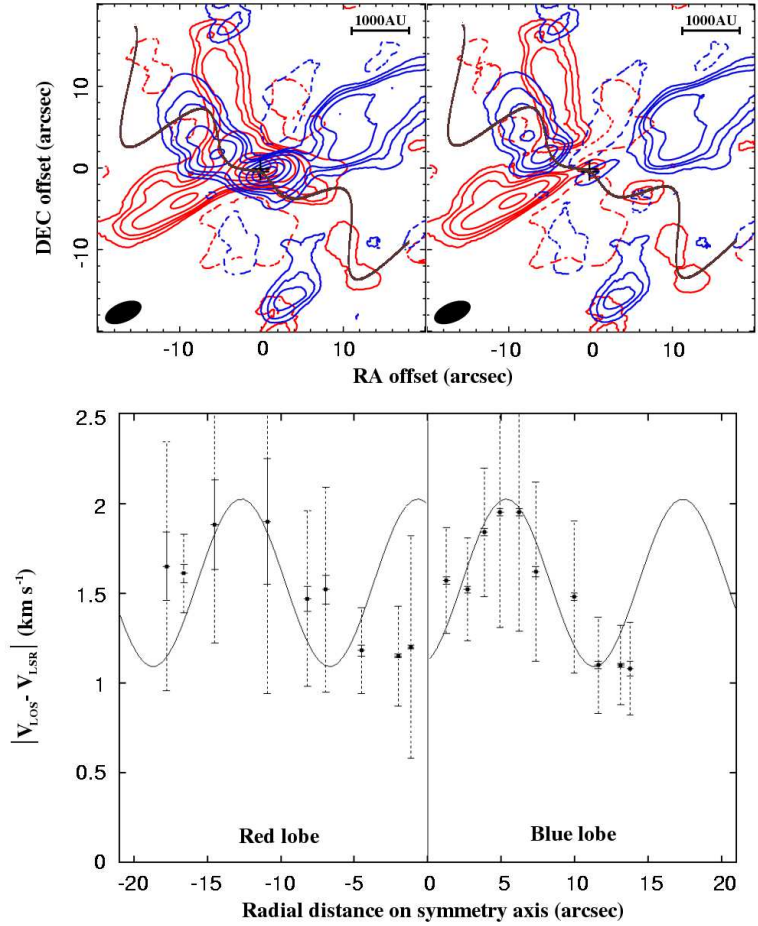}
\end{center}
\caption{Same as for Figure~11, but in this case for an outflow driven by a mirror-symmetric jet with model parameters as described in $\S4.3.3$.}
\end{figure}


\begin{thebibliography}{}

\bibitem[Andr\'e et al.(1999)]{and99} Andr\'e. P., Motte, F., Bacmann, A. 1999, \apjl, 513, 57

\bibitem[Arce \& Sargent(2004)]{arc04} Arce, H. G. \& Sargent, A. I. 2004, \apj, 612, 342

\bibitem[Arce \& Sargent(2005)]{arc05} Arce, H. G. \& Sargent, A. I. 2005, \apj, 624, 232

\bibitem[Arce \& Sargent(2006)]{arc06} Arce, H. G. \& Sargent, A. I. 2006, \apj, 646, 1070
\bibitem[Duquennoy \& Mayor(1991)]{duq91} Duquennoy, A., \& Mayor, M. 1991, \aap, 248, 485
\bibitem[Bachiller et al.(1994)]{bac94} Bachillar, R., Tafalla, M., Cernicharo, J. 1994, \apjl, 425, 93
\bibitem[Bachiller et al.(2001)]{bac01} Bachillar, R., P\'erez Guti\'errez, M., Kumar, M. S., Tafalla, M. 2001, \aap, 372, 899
\bibitem[Bally et al.(2003)]{bal03} Bally, J., Feigelson, E., Reipurth, B. 2003, \apj, 584, 843
\bibitem[Barsony et al.(1993)]{bar93} Barsony, M., Scoville, N. Z., \& Chandler, C. J. 1993, \apj, 409, 275

\bibitem[Bate et al.(2000)]{bat00} Bate, M. R., bonnell, I. A., Clarke, C. J., Lubow, S. H., Ogilvie, G. I., Pringle, J. E., Tout, C. A. 2000, \mnras, 317 733
\bibitem[Beckwith \& Sargent(1991)]{bec91} Beckwith, S. V. W. \& Sargent, A. I. 1991 \apj, 381, 250
\bibitem[Bieging \& Cohen(1985)]{bie85} Bieging, J. H., Cohen, M. 1985, \apjl, 292, 249
\bibitem[Bourke et al.(1997)]{bou97} Bourke, T. L., Garay, G., Lehtinen, K. K., Koehnenkamp, I., Launhardt, R., Nyman, L., May, J., Robinson, G., Hyland, A. R. 1997, \apj, 476, 781
\bibitem[Eisl\"offel et al.(1996)]{eis96} Eisl\"offel, J., Smith, M. D., Davis, C. J., Ray, T. P. 1996, \aj, 112, 2086
\bibitem[Fendt \& Zinnecker(1998)]{fen98} Fendt, C. \& Zinnecker, H. 1998, \aap, 334, 750
\bibitem[Fridlund et al.(1989)]{fri89} Fridlund, C. V. M., Sandqvist, A., Nordh, H. L., Olofsson, G. 1989, \aap, 213, 310
\bibitem[Fridlund \& Liseau(1998)]{fri98} Fridlund, C. V. M., Liseau, R. 1998, \apjl, 499, 75
\bibitem[Fridlund et al.(2002)]{fri02} Fridlund, C. V. M., Bergman, P., White, G. J., Pilbratt, G. L., Tauber, J. A. 2002, \aap, 382, 573
\bibitem[Gueth \& Guilloteau(1999)]{gue99} Gueth, F. \& Guilloteau, S. 1999 \aap, 343, 571

\bibitem[Ho et al.(2004)]{ho04} Ho, P. T. P., Moran, J. M., Lo, K. Y. 2004, \apjl, 616, 1 
\bibitem[Hodapp(1994)]{hod94} Hodapp, K. 1994, \apjs, 94, 615
\bibitem[J\o rgensen et al.(2007)]{jor07} J\o gensen, J. K., Bourke, T. L., Myers, P. C., Di Francesco, J., van Dishoeck, E. F., Lee, C., Ohashi, N., Sch$:o$ierm F. L., Takakuwa, S., Wilner, D. J., Zhang, Q. 2007, \apj, 659, 479
\bibitem[Keene \& Masson(1990)]{kee90} Keene, J., Masson, C. R. 1990, \apj, 355, 635  
\bibitem[Lim \& Takakuwa(2006)]{lim06} Lim, J. \& Takakuwa, S. 2006, \apj, 653, 425
\bibitem[Looney, Mundy \& Welch(1997)]{loo97} Looney, L. W. \& Mundy, L. G., \& Welch, W. J. 1997, \apjl, 484, 157
\bibitem[Masciadri \& Raga(2002)]{mas02} Masciadri, E. \& Raga, A. C. 2002, \apj, 568, 733
\bibitem[Mathieu(1994)]{mat94} Mathieu, R. D. 1994, AR\aap, 32, 465
\bibitem[Momose et al.(1998)]{mom98} Momose, M., Ohashi, N.,
Kawabe. R., Nakano, T.,\& Hayashi, M. 1998 \apj, 504, 314

\bibitem[Moriarty-Schieven et al.(1987)]{mo87} Moriarty-Schieven, G. H., Snell, R. L., Strom, S. E., Schloerb, F. P., Strom, K. M., Grasdalen, G. L. 1987, \apjl, 371, 95

\bibitem[Moriarty-Schieven \& Snell (1988)]{mo88} Moriarty-Schieven, G. H., Snell, R. L. 1988, \apj, 332, 364

\bibitem[Moriarty-Schieven et al.(1994)]{mo94} Moriarty-Schieven, G. H., Wannier, P. G., Keene, J., Tamura, M. 1994, \apj, 436, 800

\bibitem[Moriarty-Schieven et al.(2006)]{mo06} Moriarty-Schieven, G. H., Johnstone, D., Bally, J., \& Jenness, T. 2006, \apj, 645, 357

\bibitem[Ohashi et al.(1996)]{oha96} Ohashi, N., Hayashi, M., Ho, P. T. P., Momose, M., Hirano, N. 1996, \apj, 466, 957

\bibitem[Ohashi et al.(1991)]{oha91} Ohashi, N., Kawabe, R., Ishiguro, M., Hayashi, M. 1991, \aj, 102, 2054
\bibitem[Pyo et al.(2002)]{pyo02} Pyo, T., Hayashi, M., Kobayashi, N., Terada, H., Goto, M., Yamashita, T. 2002, \apj, 570, 724
\bibitem[Pyo et al.(2005)]{pyo05} Pyo, T., Hayashi, M., Kobayashi, N., Tokunage, A. T., Terada, H., Tsujimoto, M., Hayashi, S. S., Usuda, T., Yamashita, T., Takami, H., Takato, N., Kendachi, K. 2005, \apj, 618, 817
\bibitem[Rodr\'iguez et al.(1998)]{rod98} Rodr\'iguez, L. F., D'Alessio, P., Wilner, D. J., Ho, P. T. P., Torrelles, J. M., Curiel, S., G\'omez, Y., Lizano, S., Pedlar, A., Cant\'o, J., Rage, A. C. 1998, \nat, 395, 355
\bibitem[Rodr\'{\i}guez et al.(2003a)]{rod03a} Rodr\'{\i}guez, L. F., Curiel, S., Cant\'o, J., Loinard, L., Raga, A. C., \& Torrelles, J. M. 2003a, \apj, 421, 330 
\bibitem[Rodr\'{\i}guez et al.(2003b)]{rod03b} Rodr\'{\i}guez, L. F., Porras, A., Claussen, M. J., Curiel, S., Wilner, D. J., \& Ho, P. T. P. 2003b, \apj, 586, L137

\bibitem[Saito et al.(1996)]{sai96} Saito, M., Kawabe, R., Kitamura, Y., sunada, K. 1996, \apj, 473, 464
\bibitem[Sault et al.(1995)]{sau95} Sault, R. J., Teuben, P. J., Wright, M. C. H. 1995, ASPC, 77, 433
\bibitem[Scoville et al(1993)]{sco93} Scoville, N. Z., Carlstrom, J. E., Chandler, C. J., Phillips, J. A., Scott, S. L., Tilanus, R. P. J., Wang, Z. 1993, \pasp, 105, 1482
\bibitem[Snell et al.(1980)]{sne80} Snell, R. L., Loren, R. B., Plambeck, R. L. 1980, \apj, 239, 844

\bibitem[Stahler(1994)]{sta94} Stahler, S. W. 1994, \apj, 422, 616

\bibitem[Stojimirovi\'c et al.(2006)]{sto06} Stojimirovi\'c, I., Narayanan, G.,
Snell, R. L. \& Bally, J. 2006, ApJ, 649, 280
\bibitem[Strom et al.(1976)]{sto76} Strom, K. M., Strom, S. E. \& Vrba, F. J. 1976, \aj, 81, 320
\bibitem[Tafalla et al.(2000)]{taf00} Tafalla, M., Myers, P. C., Mardones, D., Bachiller, R. 2000, \aap, 359, 967
\bibitem[Takakuwa et al.(2004)]{tak04} Takakuwa, S., Ohashi, N., Ho, P. T. P., Qi, C., Wilner, D. J., Zhang, Q., Bourke, T. L., Hirano, N., Choi, M., Yang, J. 2004, \apjl, 616, 15
\bibitem[Tohline(2002)]{toh02} Tohline, J. E. 2002, AR\aap, 40, 349  

\bibitem[Uchida et al.(1987)]{uch87} Uchida, y., Kaifu, N., Shibata, K., Hayashi, S. S., Hasegawa, T., Hmatake, H. 1987, \pasj, 39, 907

\bibitem[Velusamy \& Langer(1998)]{vel98} Velusamy, T. \& Langer, W. D. 1998, \nat, 392, 685

\bibitem[Wilner \& Welch(1994)]{wil94} Wilner, D. J., Welch, W. J. 1994, \apj, 427, 898
\bibitem[Yen et al. (2009)]{yen} Yen, H-W., Takakuwa, S., in prep
\end{thebibliography}
\end{document}